\documentclass[usegraphicx]{mn2e}
\usepackage{longtable}
\pdfoutput=1
\usepackage{amsmath,amssymb}
\usepackage{times}
\usepackage{subfigure}
\usepackage{epsfig}
\usepackage{graphics}

\renewcommand{\descriptionlabel}[1]%
  {\hspace{\labelsep}\textbf{#1}}

\title[Variable stars in NGC 6229]
     {Revisiting the variable star population in NGC~6229 and the
structure of the Horizontal Branch\thanks{Based on
observations collected with the 2.0-m telescope at the Indian Astronomical
Observatory.}}

\author[A. Arellano Ferro et al.]
{A. Arellano Ferro$^{1}$,
P. E. Mancera Pi\~na$^{1,2}$,
D. M. Bramich$^{3}$,
Sunetra Giridhar$^{4}$,
\and
J. A. Ahumada$^{5}$,
N. Kains$^{6}$,
K. Kuppuswamy$^{4}$\\
$^{1}$Instituto de Astronom\1a, Universidad Nacional Aut\'onoma de M\'exico.
Ciudad Universitaria CP 04510, Mexico: (armando@astro.unam.mx)\\
$^{2}$Facultad de F\1sica, Universidad Veracruzana, Xalapa, Mexico:
(pavman\_93@hotmail.com)\\
$^{3}$Qatar Environment and Energy Research Institute, Qatar Foundation, P.O. Box
5825, Doha, Qatar: (dan.bramich@hotmail.co.uk)\\
$^{4}$Indian Institute of Astrophysics, Koramangala 560034, Bangalore, India:
(giridhar@iiap.res.in; kuppuswamy@iiap.res.in)\\
$^{5}$Observatorio Astron\'omico, Universidad Nacional de
C\'ordoba, Laprida 854, 5000 C\'ordoba, Argentina: (javier@oac.uncor.edu)\\
$^{6}$Space Telescope Science Institute, 3700 San Martin Drive,
Baltimore, MD 21218, United States of America
(nkains@stsci.edu) \\
}

\begin{document} 

\date{Accepted -----. Received ----; in original form 2015 February 12}

\pagerange{\pageref{1}--\pageref{15}} \pubyear{2014}

\maketitle 

\label{firstpage}

\begin{abstract}
We report an analysis of new $V$ and $I$ CCD time-series photometry of the distant
globular cluster NGC 6229. The principal aims were to explore the field of the cluster
in search of new variables, and to Fourier decompose the RR Lyrae light curves in
pursuit of physical parameters.We  found 25 new variables: 10 RRab, 5 RRc, 6 SR, 1 CW,
1 SX Phe, and two that we were unable to classify. Secular period changes were
detected and measured in some favourable cases. The classifications of some of the
known variables were rectified.

The Fourier decomposition of RRab and RRc light curves  was used to independently
estimate the mean cluster value of [Fe/H] and distance. From the RRab stars we found
[Fe/H]$_{\rm UVES}$=$-1.31 \pm 0.01{\rm(statistical)} \pm 0.12{\rm(systematic)}$ 
([Fe/H]$_{\rm ZW}=-1.42$),and a distance of $30.0\pm 1.5$ kpc, and from the RRc stars
we found [Fe/H]$_{\rm UVES}$=$-1.29\pm 0.12$ and a distance of $30.7\pm 1.1$ kpc,
respectively. Absolute magnitudes, radii and masses are also reported for individual
RR Lyrae stars. Also discussed are the independent estimates of the cluster distance
from the tip of the RGB, 34.9$\pm$2.4 kpc and from the P-L relation of SX Phe stars,
28.9$\pm$2.2 kpc.  The distribution of RR Lyrae stars in the horizontal branch shows a
clear empirical border between stable fundamental and first overtone pulsators which
has been noted in several other clusters; we interpret it as the red edge of the
first overtone instability strip. 
\end{abstract}


\section{Introduction}

As part of a long-term programme of CCD time-series photometry of globular
clusters (GC) aimed at
updating their variable star census and offering independent estimates of their
metallicity and distance, in the present paper we carry out the study of the
globular cluster NGC~6229
(C1645+476 in the IAU nomenclature) ($\alpha = 16^{\mbox{\scriptsize h}}
46^{\mbox{\scriptsize m}} 58.6^{\mbox{\scriptsize s}}$, $\delta = +47\degr 31\arcmin
36.4\arcsec$, J2000; $l = 73.64\degr$, $b = +40.31\degr$). This is a distant
outer-halo cluster ($d \sim$ 30 kpc) which is moderately concentrated and hence
variable stars
have been identified in regions very near its core since very early in
photographic studies of the cluster.
The Catalogue of Variable Stars in Globular Clusters (CVSGC; Clement et al. 2001;
2013 edition) lists 48 variables, mostly of the RR Lyrae type (RRL). 
The first variable star in NGC~6229, a type II cepheid, was announced by Davis (1917)
and later named as V8 by Baade (1945) who, using 46 photographic plates obtained at 
Mount Wilson in the years 1932-1935,
discovered another twenty variables, V1-V7, V9-V21. Sawyer (1953) added the W
Virginis or CW star V22
indicating that its likely period is longer than one day.
Nearly forty years passed before two more variables were found by Carney et al.
(1991), already with a CCD detector, and later confirmed by Borissova et al. (2001)
(BCV01); 
these are the two RRc stars V23 and V24. Spassova \& Borissova (1996) suggested 12
possible variables, only seven of which were confirmed by BCV01; nevertheless,
these authors numbered all 12 stars as V25-36. However, the variability status 
and/or type of stars V25, V26, V27,
V28, V29 and V30 have remained dubious in the CVSGC. In their paper,
BCV01 also reported 12 new variables, V37-48.

In most contemporary variable star studies of NGC~6229 using CCD photometry, the
recipes based on PSF or aperture photometry (Stetson 1987) and  image subtraction
algorithms such as {\tt ISIS} (Alard 2000) have been used. In the present paper we
employ the
{\tt DanDIA} implementation of difference image analysis (DIA) 
to extract high-precision photometry for all of the point
sources in the field of NGC~6229. 
In the 13 years elapsed since the last CCD time-series analysis of the cluster
(BCV01),
important improvements have taken place in the concept of DIA, which are
described in detail by Bramich (2008) and Bramich et
al. (2013). Hence, we considered that a new time-series study may reveal a
useful update of the variable star population in NGC~6229.

In previous papers we have demonstrated
how DIA performs exceptionally well in finding new variables and 
low signal-to-noise effects in light curves such as amplitude modulations in RR Lyrae
stars, 
even in the crowded central regions of globular clusters (e.g. L\'azaro et al.
2006; Bramich et al.
2011; Arellano Ferro et al. 2011, 2013
and references therein). In the present paper we report the results of the analysis
of new time-series photometry in the $V$ and $I$ filters of NGC~6229. We have 
Fourier decomposed the RR Lyrae (RRL) light curves to calculate [Fe/H], distance,
effective
temperature, mass and radius for individual stars. Mean [Fe/H] and distance are  
hence good estimates of the parent globular cluster values.

The plan of the paper is as follows. In $\S$ 2 we describe the observations and data
reduction. In $\S$ 3 the
approaches to new variable star identification are described and we report the new
variables found and discuss the secular period changes calculated in some
favourable cases. $\S$ 4 contains the
Fourier light
curve decomposition of some of the RR Lyrae stars and we report their
corresponding
individual physical parameters. $\S$ 5 is dedicated to the discussion of
the distribution of RRL stars in the HB in NGC 6229 and in other recently
studied clusters. In $\S$ 6 the metallicity and distance of the parent cluster
are inferred from the RRL stars. Independent distance calculations are also carried
out from the P-L relation of SX Phe stars, and the luminosity
of the tip of the red giant branch (RGB). In $\S$ 7 we summarise our conclusions.
Finally, the classification and peculiarities of individual stars are discussed
in Appendix A.

\section{Observations and Reductions}
\label{sec:Observations}

\subsection{Observations}

The observations were performed on 20 nights between 7th April, 2010 and  19th June,
2013 with the 2.0~m telescope at the Indian Astronomical Observatory (IAO), Hanle,
India,
located at 4500~m above sea level. A total of 346 and 354 images were obtained  in
the Johnson-Kron-Cousins $V$ and $I$ filters, respectively. The detector was a
Thompson CCD of 2048$\times$2048
pixels with a scale of 0.296 arcsec/pix, translating to a field of view (FoV) of
approximately
10.1$\times$10.1~arcmin$^2$. 

The log of observations is given in Table \ref{tab:observations} where the dates,
number of frames, exposure times and average nightly seeing are recorded. 

\begin{table}
\caption{The distribution of observations of NGC 6229 for each filter.
Columns $N_{V}$ and $N_{I}$ give the number of images taken with the $V$ and $I$
filters respectively. Columns $t_{V}$ and $t_{I}$ provide the exposure time,
or range of exposure times,
employed during each night for each filter. The 
average seeing is listed in the last column.}
\centering
\begin{tabular}{lccccc}
\hline
Date          &  $N_{V}$ & $t_{V}$ (s) & $N_{I}$ &$t_{I}$ (s)&Avg seeing (") \\
\hline
20100407 & 14            & 500          & 13            &100 &2.2\\
20100505 &  2             & 500         & 3            &100 &1.9\\
20110411 &  6             & 500-600     & 8          &100-200 &2.7\\
20110412 &  6             & 400-500      & 8          &100-200 &2.1\\
20110413 &  11             & 350-500     & 12          &75-150 &1.8\\
20110414 &  15             & 350-400      & 16     &70-80 &1.9\\
20110609 &  2             & 600         & 1             &200 &1.7\\
20110611 &  16             & 450        & 15             &120-200 &1.8\\
20110612 &  25             & 110-450         & 24             &110-120 &1.7\\
20110709 &  13             & 140-600     & 12      &140-160 &2.4\\
20110710 &  14             & 400-600    & 15      &80-120 &2.1\\
20120428 &  1             & 600         & 1            &200&2.6\\
20120429 &  5           & 600            & 1         &200&2.1\\
20130304 & 54           & 30-220          & 54         &12-80 &2.5\\
20130305 &  45           &40-100           & 47         &15-45 &2.5\\
20130306 &  41            & 50-70         & 40         &18-30 &2.3\\
20130416 & 25          & 240-400          & 28      &30-150 &2.5\\
20130417 &  19           & 400            & 20        &125-150 &2.3\\
20130618 &  18            & 300         & 20       &80-100 &1.7\\
20130619 &  14              & 300       & 16      &100 &1.7\\

\hline
Total:   & 346    &        &  354    &           &\\
\hline
\end{tabular}
\label{tab:observations}
\end{table}

\subsection{Difference Image Analysis}

We employed the technique of difference image analysis (DIA) to extract high-precision
photometry for all of the point sources in the images of NGC~6229 and we used the 
{\tt DanDIA}\footnote{{\tt DanDIA} is built from the DanIDL library of IDL routines
available at http://www.danidl.co.uk}
pipeline for the data reduction process (Bramich et al. 2013). {\tt DanDIA}
includes an 
algorithm that models the convolution kernel matching the PSF
of a pair of images of the same field as a discrete pixel array (Bramich 2008).
The {\tt DanDIA} pipeline performs standard bias level and flat field
corrections
of the raw images, and creates a reference image for each filter by stacking a set of
registered best-seeing calibrated images.

\begin{figure} 
\includegraphics[scale=0.4]{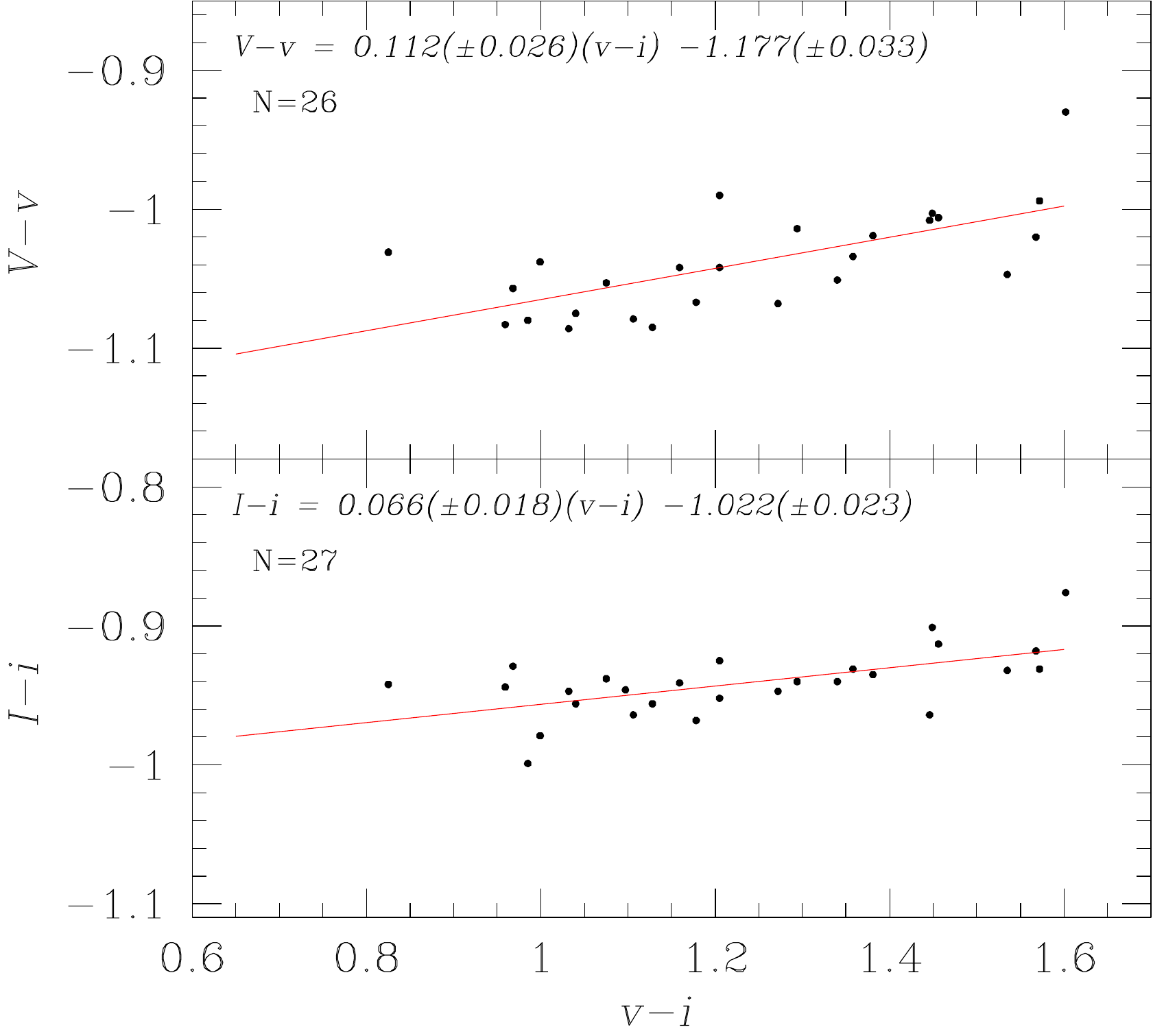}
\caption{Transformation relations in the $V$ and $I$ band-passes between the
instrumental and the standard
photometric systems using a set of standard stars in the field of NGC~6229
kindly provided by Peter Stetson. }
    \label{transV}
\end{figure}

We constructed two reference images, one in
the $V$ filter and another in $I$. For each of these
reference images, 5 calibrated images were stacked with total exposure times of
1800 and 370 seconds in the $V$ and $I$  filters, respectively, and
with PSF FWHMs of $\sim$5.2 and $\sim$4.0~pix, respectively.
In each reference image, we measured the fluxes (referred to as reference fluxes) 
and positions of all PSF-like objects (stars)
by extracting a spatially variable (with a third-degree polynomial) empirical PSF from
the 
image and fitting this PSF to each detected object. The detected stars in each image
in the time-series were matched with those detected in the corresponding reference
image, and a linear transformation was derived which was used to register each image
with the reference image.

\begin{table*}
\caption{Time-series $V$ and $I$ photometry for all the confirmed variables in our 
field of view. Filter and epoch of
mid-exposure are listed in columns 2 and 3, respectively. The standard
$M_{\mbox{\scriptsize std}}$ and
instrumental $m_{\mbox{\scriptsize ins}}$ magnitudes are listed in columns 4 and 5,
respectively, corresponding to the variable star in column 1.  The uncertainty on
$m_{\mbox{\scriptsize ins}}$ is listed in column 6, which also corresponds to the
uncertainty on $M_{\mbox{\scriptsize std}}$. For completeness, we also list the
quantities
$f_{\mbox{\scriptsize ref}}$, $f_{\mbox{\scriptsize diff}}$ and $p$ from
Equation~\ref{eqn:totflux}
in columns 7, 9 and 11, along with the uncertainties $\sigma_{\mbox{\scriptsize
ref}}$ and $\sigma_{\mbox{\scriptsize diff}}$ in columns 8 and 10. This is an extract
from the full table, which is available with the electronic version of the article
(see Supporting Information).}
\centering
\begin{tabular}{ccccccccccc}
\hline
Variable &Filter & HJD & $M_{\mbox{\scriptsize std}}$ &
$m_{\mbox{\scriptsize ins}}$
& $\sigma_{m}$ & $f_{\mbox{\scriptsize ref}}$ & $\sigma_{\mbox{\scriptsize ref}}$ &
$f_{\mbox{\scriptsize diff}}$ &
$\sigma_{\mbox{\scriptsize diff}}$ & $p$ \\
Star ID  &        & (d) & (mag)                        & (mag)                       
& (mag)        & (ADU s$^{-1}$)               & (ADU s$^{-1}$)                    &
(ADU s$^{-1}$)                &
(ADU s$^{-1}$)                     &     \\
\hline
V1& V & 2455294.32981&17.971&19.087& 0.009& 179.343& 0.957& +26.504 &0.931& 0.5048\\
V1& V & 2455294.34033&18.017&19.133& 0.007& 179.343& 0.957& +26.841 &0.923& 0.6244\\
\vdots   & \vdots & \vdots  & \vdots & \vdots & \vdots & \vdots   & \vdots & \vdots &
\vdots & \vdots \\
V1& I & 2455294.31704&17.435& 18.422& 0.029&328.424& 2.483& +44.876& 4.482 &0.4514\\
V1& I & 2455294.32313&17.432& 18.418& 0.023&328.424& 2.483& +52.696& 4.111 &0.5226\\
\vdots   & \vdots & \vdots  & \vdots & \vdots & \vdots & \vdots   & \vdots & \vdots  &
\vdots & \vdots \\
V2& V & 2455294.32981&18.140& 19.251& 0.011& 194.212& 0.956& +2.550 &0.944&0.5048\\
V2& V & 2455294.34033&18.147& 19.259& 0.009& 194.212& 0.956& +2.271 &0.942&0.6244\\
\vdots   & \vdots & \vdots  & \vdots & \vdots & \vdots & \vdots   & \vdots & \vdots  &
\vdots & \vdots \\
V2& I & 2455294.31704&17.530&18.514&0.032& 336.249&3.188& +25.571 &4.560&0.4514\\
V2& I & 2455294.32313& 17.552&18.536&0.026& 336.249& 3.188& +25.545&4.237&0.5226\\
\vdots   & \vdots & \vdots  & \vdots & \vdots & \vdots & \vdots   & \vdots & \vdots  &
\vdots & \vdots \\
\hline
\end{tabular}
\label{tab:vi_phot}
\end{table*}

\begin{figure}
\includegraphics[width=8.0cm,height=11.5cm]{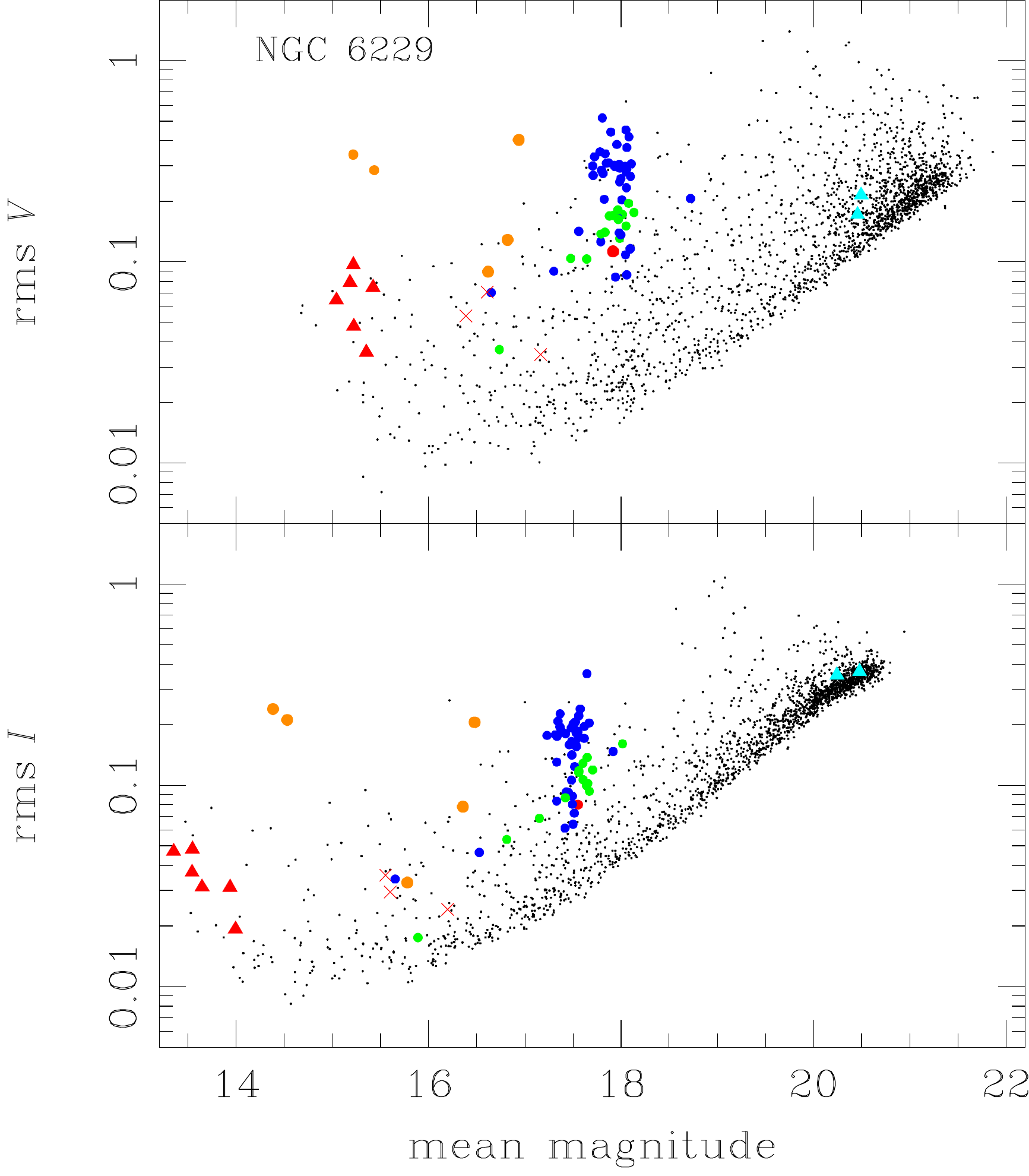}
\caption{The rms magnitude deviations as a function of
the mean magnitudes $V$ and $I$. Blue and green circles are used for RRab and RRc
stars respectively.  
Other symbols are: red circle, V36 reclassified as RRc in this paper; orange
circles, CWA, CWB or CW?; red triangles, SR; turquoise triangles, variable blue
stragglers; red crosses,
previously suspected variables not confirmed in this paper.}
\label{rms}
\end{figure}

\begin{figure}
\includegraphics[width=8.0cm,height=11.5cm]{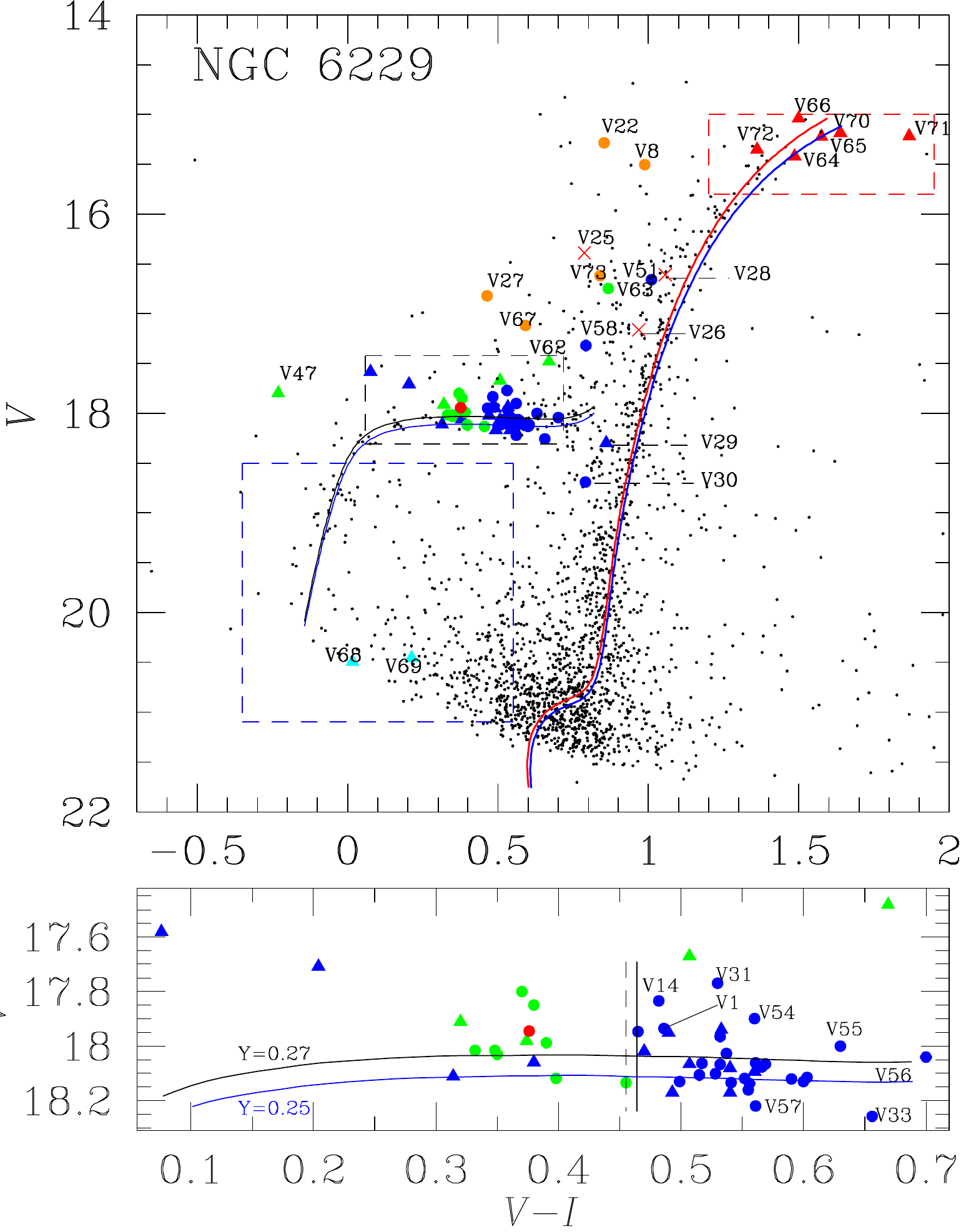}
\caption{The top panel shows the colour-magnitude
diagram of NGC~6229. Symbols and colours are as in Fig. \ref{rms}.
All variables labelled with numbers larger
than 48 are new variables announced in the present work. See the
text for a discussion on individual stars. The three dashed boxes are regions where
individual star light curves were explored for variability. The isochrones are from
the Victoria-Regina stellar models (VandenBerg et al. 2014) for an age of 12 Gyr
and [Fe/H]=$-1.42$ (red) and $-1.31$ (blue) respectively.
The bottom panel is a
blow-up of the HB region, where triangles represent stars with amplitude
modulations and the two vertical black lines are two empirical
loci of the border between RRab and RRc distributions seen in NGC~5024 (solid) and
NGC~4590 (dashed) after considering the corresponding reddenings (see the
discussion in $\S$ \ref{sec:HB}). Two ZAHB models calculated from the
Victoria-Regina stellar models for [Fe/H]=$-1.42$, [$\alpha$/Fe]=+0.4 and two values
of helium content $Y$ 0.25 and 0.27 are included (see the discussion in $\S$ 5).}
\label{CMD}
\end{figure}

For each filter, a sequence of difference images was created by 
subtracting the relevant reference image, convolved with an appropriate spatially
variable kernel, from each registered image. The spatially variable convolution
kernel for each registered image was determined using bilinear interpolation of a
set of kernels that were derived for a uniform 6$\times$6 grid of subregions across
the
image.

The differential fluxes for each star
detected in the reference image were measured on each difference image. 
Light curves for each star were constructed by calculating the total 
flux $f_{\mbox{\scriptsize tot}}(t)$ in ADU/s at each epoch $t$ from:
\begin{equation}
f_{\mbox{\scriptsize tot}}(t) = f_{\mbox{\scriptsize ref}} +
\frac{f_{\mbox{\scriptsize diff}}(t)}{p(t)},
\label{eqn:totflux}
\end{equation}
where $f_{\mbox{\scriptsize ref}}$ is the reference flux (ADU/s),
$f_{\mbox{\scriptsize diff}}(t)$ is the differential flux (ADU/s), and
$p(t)$ is the photometric scale factor (the integral of the kernel solution).
Conversion to instrumental magnitudes was achieved using:
\begin{equation}
m_{\mbox{\scriptsize ins}}(t) = 25.0 - 2.5 \log \left[ f_{\mbox{\scriptsize tot}}(t)
\right],
\label{eqn:mag}
\end{equation}
where $m_{\mbox{\scriptsize ins}}(t)$ is the instrumental magnitude of the star 
at time $t$. Uncertainties were propagated in the correct analytical fashion.

The above procedure and its caveats have been described in detail in Bramich et al.
(2011), to which the interested reader is referred for further details.

\subsection{Photometric Calibrations}

\subsubsection{Relative calibration}
\label{sec:rel}

All photometric data suffer from systematic errors to some level that sometimes may be
severe enough
to be mistaken for bona fide variability in light curves. However,
multiple observations of a set of objects at different epochs, such as time-series
photometry,
may be used to investigate, and possibly correct, these systematic errors (see for
example
Honeycutt 1992). This process is a relative self-calibration of the photometry, which
is being
performed as a standard post-processing step for large-scale surveys (e.g. Padmanabhan
et al. 2008;
Regnault et al. 2009).

We apply the methodology developed in Bramich \& Freudling (2012) to solve for the
magnitude offsets
$Z_{k}$ that should be applied to each photometric measurement from the image $k$. In
terms of DIA,
this translates into a correction (to first order) for the systematic error introduced
into the photometry
from an image due to an error in the fitted value of the photometric scale factor $p$.
We found that, for either filter, the magnitude offsets that we derive are of the
order of $\sim0.04$ mag with $\sim10$\% of worse cases reaching $\sim0.2$ mag.
Applying
these magnitude offsets
to our DIA photometry improves the light curve quality, especially for the
brighter stars.

\subsubsection{Absolute calibration}
\label{absolute}

Standard stars in the field of NGC~6229 were not included in the online collection of
Stetson
(2000)\footnote{
http://www3.cadc-ccda.hia-iha.nrc-cnrc.gc.ca/community/STETSON/standards} at the time 
of preparing this paper.
However, Prof. Stetson kindly provided us with a set of preliminary standard 
stars which we have used to transform instrumental $vi$ magnitudes into the standard
\emph{VI} system. 

The standard minus the instrumental magnitudes show mild dependences on the colour,
as can be seen in Fig.\ref{transV}. The transformations are of the form: 

\begin{equation}
V_{std}= v +0.112(\pm0.026)(v-i) - 1.177(\pm0.033),
\label{eq:transV}
\end{equation}
\begin{equation}
I_{std}= i +0.066(\pm0.018)(v-i) - 1.022(\pm0.023).
\label{eq:transI}
\end{equation}

All of our \emph{VI} photometry for the variable stars in the FoV of our
collection
of images
of NGC~6229 is provided in Table \ref{tab:vi_phot}. Just a small portion of this
table is given in the printed version of this paper, while the full table is
only available in electronic form.

Fig. \ref{rms} shows the rms magnitude deviation in our $V$ and $I$ light curves, 
after the application of the relative photometric calibration of
Section~\ref{sec:rel}, as a function of the mean magnitude. 

To help us discuss the variable star search and classifications, we have
built 
the colour-magnitude diagram (CMD) of Fig. \ref{CMD} by calculating the
inverse-variance weighted mean magnitudes
of ∼2113 stars with $V$ and $I$ magnitudes. For better precision, the periodic
variables like RRL and SX Phe are plotted using their intensity-weighted
mean magnitudes $<V>$ and colours $<V>-<I>$. The use of $<V-I>$ instead does 
not produce
a significant difference in the plotting on the CMD as they differ from $<V>-<I>$
within 0.01 mag. The regions marked on the plot,
individual stars,
and the HB inset will be discussed below in the corresponding sections. 

\begin{figure} 
\includegraphics[width=8.0cm,height=8.0cm]{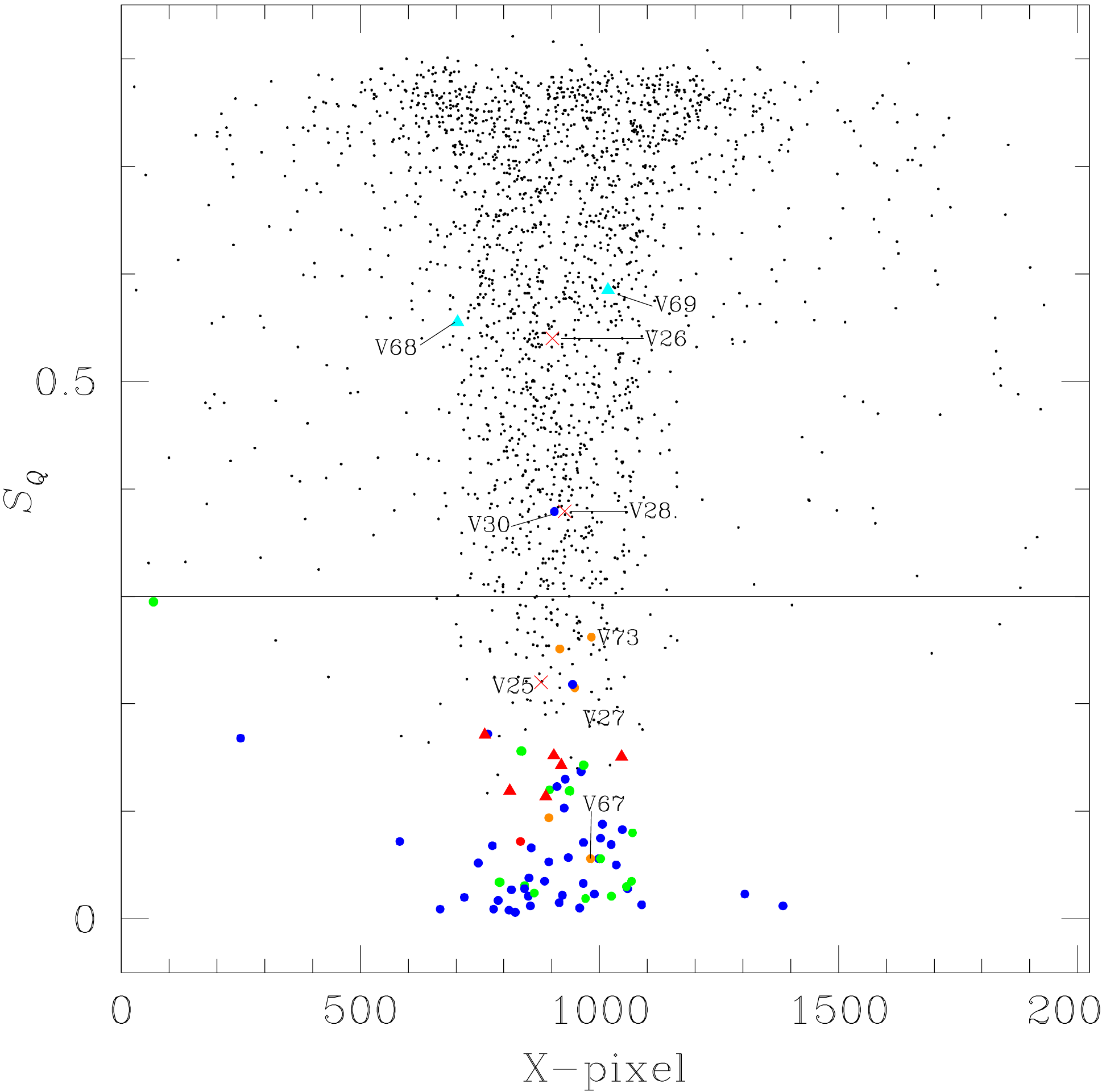}
\caption{Minimum value of the string-length parameter $S_Q$ calculated for the 2113
stars with a light curve in our $V$ reference image, 
versus the CCD $X$-coordinate. Colours are as in Fig. \ref{rms}. The horizontal
line
is an arbitrarily defined threshold
below which most variables seem to fall. All stars below the threshold were
individually explored for variability. The only true variables above the threshold
are one SX Phe (V68), one unclassified variable blue straggler (V69),
and one noisy RRab (V30).}
\label{SQ}
\end{figure}

\subsection{Astrometry}
\label{sec:astrometry}

A linear astrometric solution was derived for the
$V$ filter reference image by matching 84 hand-picked
stars with the UCAC4 star catalogue (Zacharias et al. 2013)
using a field overlay in the image display tool
{\tt GAIA}\footnote{http://star-www.dur.ac.uk/\~{}pdraper/gaia/gaia.html}
(Draper 2000). We achieved a radial RMS scatter in the residuals of
$\sim$0.2~arcsec. The astrometric fit was then used to calculate the
J2000.0 celestial coordinates for all of the confirmed variables in our field of
view (see Table~\ref{variables}). The coordinates correspond to the epoch of the $V$
reference image which pertains to the average heliocentric Julian day of the five
images used to form the reference image, 2455665.92~d.

\begin{table*}
\caption{General data for all of the confirmed variables in NGC~6229 in the FoV of 
our images. Stars V49-V73 are new discoveries in this work. Label 'Bl'
is used for Blazhko variables. The number of digits quoted for $<V>$ and $<I>$ depend
on the scatter and/or amplitude modulations seen in each light curve.}
\label{variables}
\centering
\begin{tabular}{lllllllclcll}
\hline
Variable & Variable & $<V>$   & $<I>$   & $A_V$       & $A_I$   & $P$ (days)    &
$\beta$      & HJD$_{\rm max}$     & $P$ (days)     & RA          & Dec.         \\
Star ID  & Type     & (mag)   & (mag)   & (mag)       & (mag)   & this work     
& (d Myr$^{-1}$) & (+2~450~000)    & (BCV01)  & (J2000.0)   & (J2000.0)   
\\
\hline
V1       & RRab     & 17.936  & 17.450  & 1.16       & 0.82   & 0.585700   
 & +0.120 &6399.3193       & 0.5857 &16:46:56.34  &+47:29:53.3  \\
V2       & RRab-Bl    & 18.07 & 17.56  &1.01    & 0.63    & 0.555239   
 & +0.044  & 6357.4046      & 0.5553 & 16:46:51.62   &+47:31:43.5  \\
V3       & RRab  & 18.13 & 17.59  & 0.89          & 0.67 & 0.575259     
& --0.432 & 5753.1692       & 0.5752 &16:46:39.38 & +47:32:19.3 \\
V4       & RRab-Bl & 18.09 & 17.53  & 0.86     & 0.58 & 0.566196    
&   --0.500  & 5664.3882       & 0.5662 &16:46:53.12 & +47:31:24.3 \\
V5       & RRab      & 18.160 & 17.605  & 1.05      & 0.68  & 0.533602 &     
 --0.056 & 5725.3179       & 0.5336 &16:47:00.11  & +47:32:23.0\\
V6       & RRab-Bl     & 18.08 & 17.54  & 0.85     & 0.52   & 0.559349  &  
--  & 6462.3056       & 0.5593 &16:47:03.03  &+47:32:20.7 \\
V7       & RRab-Bl   & 18.02 & 17.55 & 1.15   & 0.77   & 0.506969  &    
 -- & 5294.3403    & 0.5016 & 16:46:54.60&+47:30:48.8  \\
V8       & CWA  &15.50 &14.62 & 0.75    & 0.58 &  14.846          
&    
 -- &  6462.3310           & 14.8405  &16:46:58.33  &+47:30:56.9 \\
V9       & RRab-Bl    & 17.95 & 17.46   & 0.86    & 0.63  & 0.542841   &   
 --0.204   & 6357.4281       & 0.5428 & 16:46:54.85 &+47:32:17.0 \\
V10      & RRab   & 18.101 & 17.533  & 1.00       & 0.65  & 0.554738   &     
 --0.280 & 6463.3279       &0.5547 &16:46:55.76  &+47:32:51.4 \\
V11      & RRab     & 18.119 & 17.566  &  0.92   & 0.59   & 0.563226   &     
--    & 5752.2791       & 0.5632  &16:47:01.08 & +47:31:14.1 \\
V12  & RRab-Bl & 17.91  & 17.67 & 1.16      & 0.95 & 0.490812    &
--          & 6463.3125    &  0.4908& 16:47:02.11 & +47:31:15.4 \\
V13      & RRab     & 18.106  & 17.591  & 0.93    & 0.62  & 0.547345   &     
 +0.056  & 6357.3728     & 0.5473 & 16:47:12.51 & +47:32:41.0 \\
V14      & RRab      & 17.840 &  17.353 & 1.04     & 0.66 & 0.465926  &
 +0.096 & 5725.3366   & 0.4659 & 16:46:57.21 & +47:30:48.1 \\
V15      & RRc     & 18.12 & 17.72 & 0.52     & 0.32  & 0.271379   &    
 +0.024 & 5722.2369  & 0.2714 & 16:47:02.07 & +47:32:06.7 \\
V16      & RRc-Bl  & 17.98 & 17.61 & 0.49       & 0.30 & 0.322836 &    
 +0.780       & 6399.2877   & 0.3227 & 16:47:03.36 & +47:31:15.0\\
V17      & RRc-Bl     & 17.91 & 17.59  & 0.46       & 0.35   & 0.325141  
& --  &6463.2279     &0.3252  & 16:46:49.25 & +47:30:23.4 \\
V18      & RRab  & 18.066 & 17.497  & 0.79      & 0.51 & 0.579039     
&--  & 5725.3594   & 0.5791& 16:46:55.13 & +47:32:11.0 \\
V19      & RRab    & 18.130 & 17.631 & 1.29       & 0.87   & 0.475960   
&  --0.024   & 6462.3457       & 0.4759 & 16:47:03.99 & +47:30:54.9 \\
V20      & RRab-Bl     & 18.17 & 17.68  & 1.30    & 0.88 & 0.465952
 & --0.232  & 6046.3548     & 0.4660        & 16:46:56.03 & +47:31:02.8 \\
V21      & RRab    &  18.063  & 17.546  & 0.98     & 0.66   & 0.564426    
&  --    & 6462.3941       & 0.5644 & 16:47:10.32 & +47:30:38.0 \\
V22      & CWA   &15.28&14.43& 0.88     & 0.72   & 15.846     &     
--       & 6047.4139   & 15.8373 & 16:46:58.95 & +47:31:28.1 \\
V23      & RRc   & 17.868 & 17.433 & 0.38      & 0.22   & 0.397688   &
 --           & 5722.2369    & 0.3977 & 16:46:56.72 & +47:32:36.0 \\
V24      & RRc      & 18.016  & 17.668 & 0.44 & 0.30   &  0.303441   
& --             & 5752.2182   & 0.3034 & 16:46:57.45 & +47:30:40.5 \\
V27      & CW?    &16.83&16.35& 0.49 &0.31&  -- &    
 --       &  6358.3593      & 1.13827 & 16:46:59.85 & +47:31:46.1\\
V29      & RRab-Bl    & 18.30  & 17.44  & 1.50    & 0.74       & 0.629509  &  
 --    &  6357.4207    & --& 16:47:00.39 & +47:31:41.5 \\
V30      & RRab      & 18.7  &17.9&0.70   &0.35  &  0.498700   &    
--        & 6356.3490    & -- & 16:46:58.63 & +47:31:23.3 \\
V31      & RRab     & 17.77  & 17.24  & 0.88   & 0.61 &   0.537851  &    
--        & 6462.3941   & 0.6989 & 16:46:59.30 & +47:31:18.9\\
V32      & RRab-Bl   & 18.17  & 17.63  & 0.61      & 0.49   & 0.603805 &   
--      &  5752.2266    & 0.3765   & 16:46:58.34 & +47:31:17.7 \\
V33      & RRab    & 18.26  & 17.60  & 1.24       & 0.78  & 0.517500    &     
--      & 6356.3640  & 0.5176 & 16:46:58.24 & +47:32:01.1\\
V34      & RRab      & 18.14 & 17.58  & 0.99     & 0.61       &  0.555409  &   
--        & 6356.3936   & 0.5554 & 16:47:01.43 & +47:31:44.7\\
V35      & RRab      & 18.062 & 17.501   & 0.31     & 0.23   & 0.634460    &     
--        & 5665.3884  &0.6345 & 16:47:01.53 & +47:31:57.2 \\
V36      & RRc      & 17.942 & 17.569  & 0.32    & 0.22 & 0.264133      &    
--        & 6399.2877  & 0.2641 & 16:46:56.51 & +47:32:02.9 \\
V37      & RRab      & 17.947 & 17.482  & 0.83     & 0.63 & 0.519174  &     
--       & 6399.4511  & 0.5192 & 16:46:58.87 & +47:32:11.9\\
V38      & RRab-Bl      & 18.06  & 17.68   & 1.06  & 0.78    & 0.522159   &  
--        &  5665.3610 &0.5222  & 16:46:57.09 & +47:31:11.4 \\
V39      & RRc      & 18.113 & 17.678  & 0.42     & 0.31  & 0.332997    &    
--        & 5725.3982   & 0.4993 & 16:47:03.05 & +47:31:08.1\\
V40      & RRab-Bl   & 17.58  & 17.51  & 0.51   & 0.50  & 0.591408  &    
--        & 6399.2877   & 0.5914 & 16:47:02.37 & +47:31:58.8 \\
V41     & RRab-Bl?     & 18.12 & 17.51  & 0.42   & 0.31   & 0.634608 &     
--      & 5725.1939  & 0.6347 & 16:47:02.78 & +47:31:30.4 \\
V42      & RRab     & 18.07  & 17.54  & 0.50      & 0.37    & 0.621730  &    
--      & 5664.4199   & 0.6218  & 16:47:01.30 & +47:32:00.9 \\
V43      & RRab-Bl  & 17.94  & 17.41  & 1.14      & 0.73   & 0.567649 &  
--        & 5725.1641    & 0.5677 & 16:46:56.99 & +47:32:01.8\\
V44      & RRc     & 17.988  & 17.598   & 0.45     & 0.33   & 0.357365  & 
--        & 6356.4737   & 0.3573 & 16:47:00.57 & +47:30:51.6 \\
V45      & RRab   & 18.027 & 17.486  & 0.48    & 0.33  & 0.640061   &     
--        &  6356.3490   & 0.6401& 16:46:59.18  &+47:30:22.2  \\
V46      & RRab     & 17.965  & 17.433    & 0.28      & 0.22 &  0.638741  & 
 --       & 6400.3493   &0.6388 & 16:46:53.97 &+47:31:24.8  \\
V47      & RRc-Bl      &  17.80 & 18.03 & 0.36  & 0.40   &   0.333948&
--        & 5724.3047   & 0.3339& 16:47:03.40 & +47:31:26.5 \\
V48      & RRc    & 17.85 & 17.47 & 0.38    & 0.26 & 0.340270 &     
--       & 5753.1385  & 0.5164 & 16:47:01.46 & +47:31:20.0 \\
V49      & RRab      & 18.08  & 17.51  & 0.35    & 0.24  & 0.637320   &    
--        &  5666.3954   & -- & 16:46:54.75 & +47:32:36.5\\
V50      & RRab-Bl     & 18.11  & 17.80  & 1.25    & 0.96  & 0.492440  &    
--        &  5752.3148  & --  & 16:46:56.78 & +47:31:43.9\\
V51      & RRab     & 16.66 & 15.65 & 0.26    & 0.11  & 0.548110  &    
--        &  5725.4078      & -- & 16:46:57.21 & +47:31:27.4 \\
V52     & RRab     & 18.13  & 17.53  &  1.05    & 0.57    & 0.491680 &     
--      & 6356.3733 & --& 16:46:57.98 & +47:32:02.3 \\
V53      & RRab-Bl?    & 18.12  & 17.53  & 0.36   & 0.25    & 0.625180   &    
--      & 5753.2588    & -- & 16:46:58.62 & +47:33:41.4 \\
V54      & RRab    & 17.90  & 17.34     & 0.73     & 0.47  & 0.570370  & 
--        & 6399.2877   & -- & 16:46:59.20 & +47:31:50.2 \\
V55      & RRab     & 18.00  & 17.37   & 1.19    & 0.64   & 0.525180   &   
--        & 6462.3893  & --  & 16:46:59.48 & +47:31:18.1 \\
V56      & RRab  & 18.04 & 17.34 & 0.41     &0.28 &  0.638310    &    
--        & 6356.3490    & -- & 16:46:59.70 & +47:31:52.6\\
V57     & RRab      & 18.22  & 17.66  & 0.99  & 0.70   &  0.561640 &     
--      & 5753.1692    & -- & 16:47:00.25 & +47:31:31.2 \\
V58      & RRab     & 17.319  & 16.527 & 0.30   & 0.16    & 0.639500 &       
--      & 5725.3496   & -- & 16:47:00.34 & +47:32:14.6 \\
V59      & RRc     & 18.0 & 17.7 & 0.55      & 0.21    & 0.335780   &     
--        & 5663.4655 & -- & 16:46:34.04 & +47:32:26.0 \\
V60      & RRc     & 18.016 & 17.684  & 0.38      & 0.25   & 0.260380  &     
--        & 6358.4103      & -- & 16:46:55.27 & +47:31:31.0\\
V61     & RRc-Bl    & 17.67  & 17.16   & 0.31     & 0.20   & 0.343050  &     
--        & 5753.2932   & -- & 16:46:56.59 & +47:31:41.6\\
V62      & RRc-Bl      & 17.48  & 16.81   & 0.27      & 0.16   &  0.299040  & 
--        & 6357.4281  & -- & 16:46:59.52 & +47:32:02.0 \\
V63      & RRc    & 16.750 & 15.893  & 0.12    & 0.05  & 0.341660   &    
--        & 5752.2898   & -- & 16:47:00.41 & +47:31:27.6 \\
\hline
\hline
\end{tabular}
\end{table*}

\begin{table*}
\addtocounter{table}{-1}
\caption{Continued}
\label{variablesB}
\centering
\begin{tabular}{lllllllclcll}
\hline
Variable & Variable & $<V>$   & $<I>$   & $A_V$       & $A_I$   & $P$ (days)    &
$\beta$      & HJD$_{max}$     & $P$ (days)     & RA          & Dec.         \\
Star ID  & Type     & (mag)   & (mag)   & (mag)       & (mag)   & this work     & (d
Myr$^{-1}$) & (+2~450~000)    & (BCV01)& (J2000.0)   & (J2000.0)    \\
\hline
V64      & SR     & 15.42  & 13.94 & 0.29 & 0.13 & --  &     
 --      & --     & -- & 16:46:55.87 & +47:31:59.9 \\
V65      & SR     & 15.22 & 13.65 & 0.17 & 0.12 & --   &    
 --       & --    & -- & 16:46:58.11 & +47:31:20.5 \\
V66      & SR     & 15.04 & 13.54 & 0.33 & 0.16  & --  &    
--        & --    & --& 16:46:59.01 & +47:31:58:0 \\
V67      & CWB      &17.12  &16.53  & 1.35 & 0.84 & 1.51547&    
--        & 5724.1700    & --& 16:47:00.83 & +47:31:41.3 \\
V68     & SX Phe      & 20.49 &20.48 &  0.4    & -- &0.0384563 &    
--      & 5663.3201     & --& 16:46:52.57 & +47:33:23.1 \\
V69      & ?    &20.46&20.24 & 0.22 &  --  & 0.271264&   
--      & 6356.3416 & -- &16:47:01.85  &+47:32:13.3 \\  
V70      & SR   & 15.18  & 13.55 & 0.36 & 0.21 &--&    
--      & --       & -- &16:46:54.22 &+47:33:37.2  \\  
V71      & SR   &15.22 & 13.35 & 0.39 & 0.20 & --&       
 --     &--        & -- & 16:46:58.58 & +47:31:36.9  \\ 
V72      & SR   & 15.35 & 13.99 & 0.12 & 0.10 &--  &  
--      & --       & -- &16:47:02.81&+47:30:21.2  \\
V73      & CW?   & 16.62 &15.78  &--  &--  &--  &  
--      & --       & -- &16:47:00.85&+47:31:43.4  \\
\hline
\end{tabular}
\end{table*}

\section{Variable stars in NGC~6229}

\subsection{Search for new variable stars}
\label{sec:IDVAR}

All previously known 48 variables were identified by their equatorial
coordinates listed in the CVSGC.
The main approach to identify new variables was the string-length
method (Burke, Rolland \& Boy 1970; Dworetsky 1983),
in which each light curve is phased with periods between
0.02 and 1.7 day and a normalised string-length statistic $S_Q$ is calculated for 
each trial period.  Extending the period base to values $~\sim2.5$d does not
produce a significantly different diagram. On the other hand if the period range is
stretched to much longer periods, i.e. $>10$d, the routine tends to find multiples of
the true periods for some RR Lyrae. The long period variables are to be spotted by
the other methods described below.
In Fig. \ref{SQ} we plot the minimum $S_Q$
value for each light curve as a function of their corresponding CCD $X$-coordinate.
The known variables are plotted with the coloured symbols described in the caption.
The horizontal line is a threshold set by eye at $S_Q=0.30$, below which most
variables seem to fall and then it is below this line where we might expect to find
previously undetected variables. 
Then we carefully inspected for variability the
light curve of each star with $S_Q$ below 0.30. This approach allowed us to
detect the new variables now labelled V49-V67 and to confirm the
non-variability of
stars V26 and V28. Despite falling below the threshold, V25 does not show
convincing signs of variability. We also note that this method did not work for 
the SX Phe star, a fact already pointed out before by Arellano Ferro et al. (2010) 
in the study of NGC~5024.

A second approach towards searching for variable stars was to separate the light
curves of stars in a
given region of the CMD where variables are expected, e.g. HB, blue stragglers
region, upper instability strip and the RGB. A further more detailed inspection
of stars in these regions revealed two new variables in the blue stragglers
region, V68 and V69. At least star 
V68 can be classified as SX Phe. Also the SR variables labelled V70-V72 and 
a possible CW star V73, which we had missed from the string-length method, were found
in this way.

Finally, we used a third approach that consisted in detecting PSF-like peaks in a
stacked image built from the sum of the 
absolute valued difference images normalised by the standard deviation in each pixel
as described by Bramich et al. (2011). This method allowed us to confirm the
variability
of all of the new variables discovered by the previous methods but no new variables
emerged. 

\begin{figure*}
 \centering
\includegraphics[scale=.4]{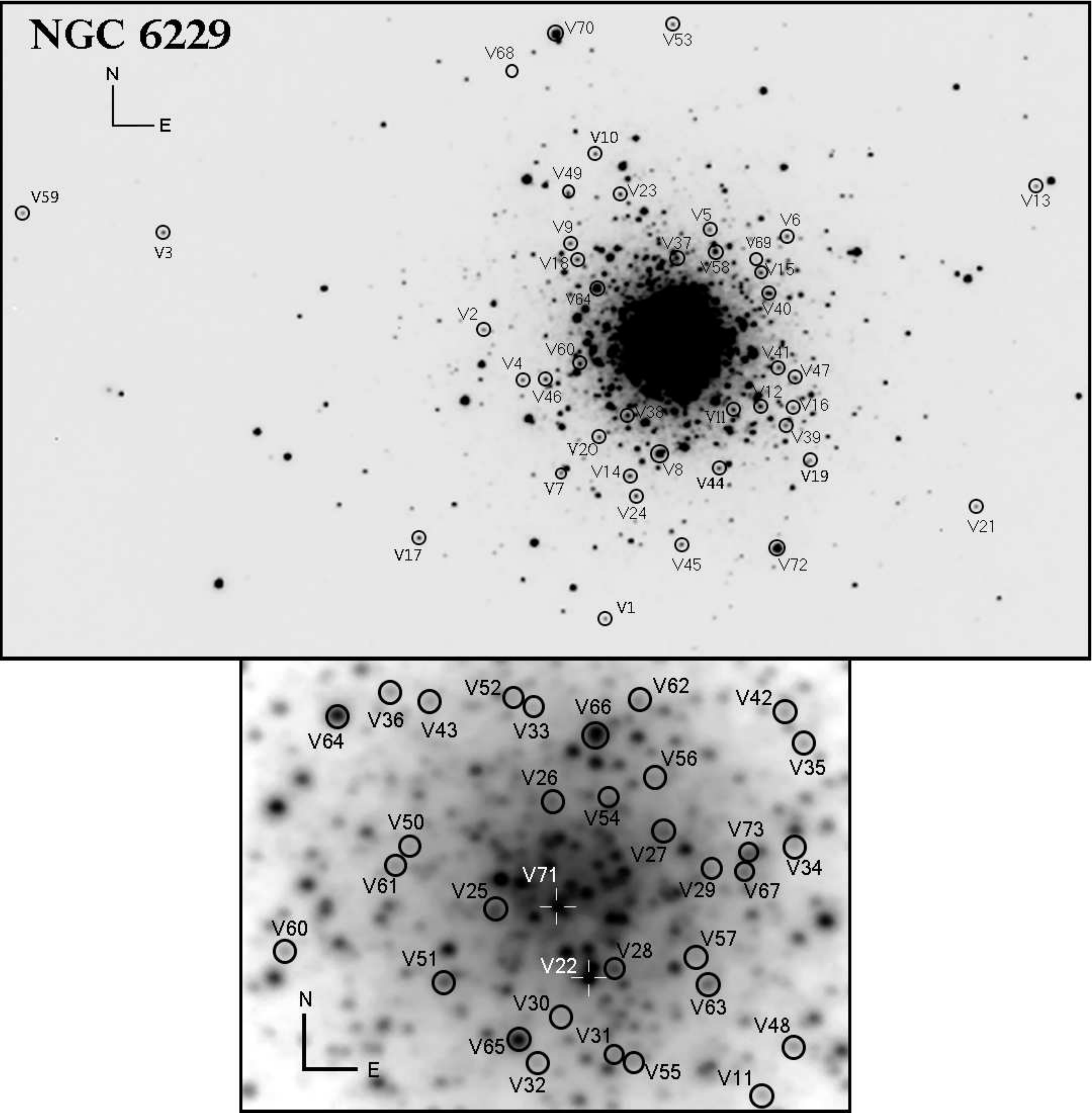}
\caption{Finding charts constructed from our $V$ reference image; north is up
and east is to the right. The cluster images are 7.3$\times$4.4 and
1.25$\times$0.90~arcmin$^{2}$.}
\label{chart}
\end{figure*}

\begin{figure*}
 \centering
\includegraphics[scale=0.85]{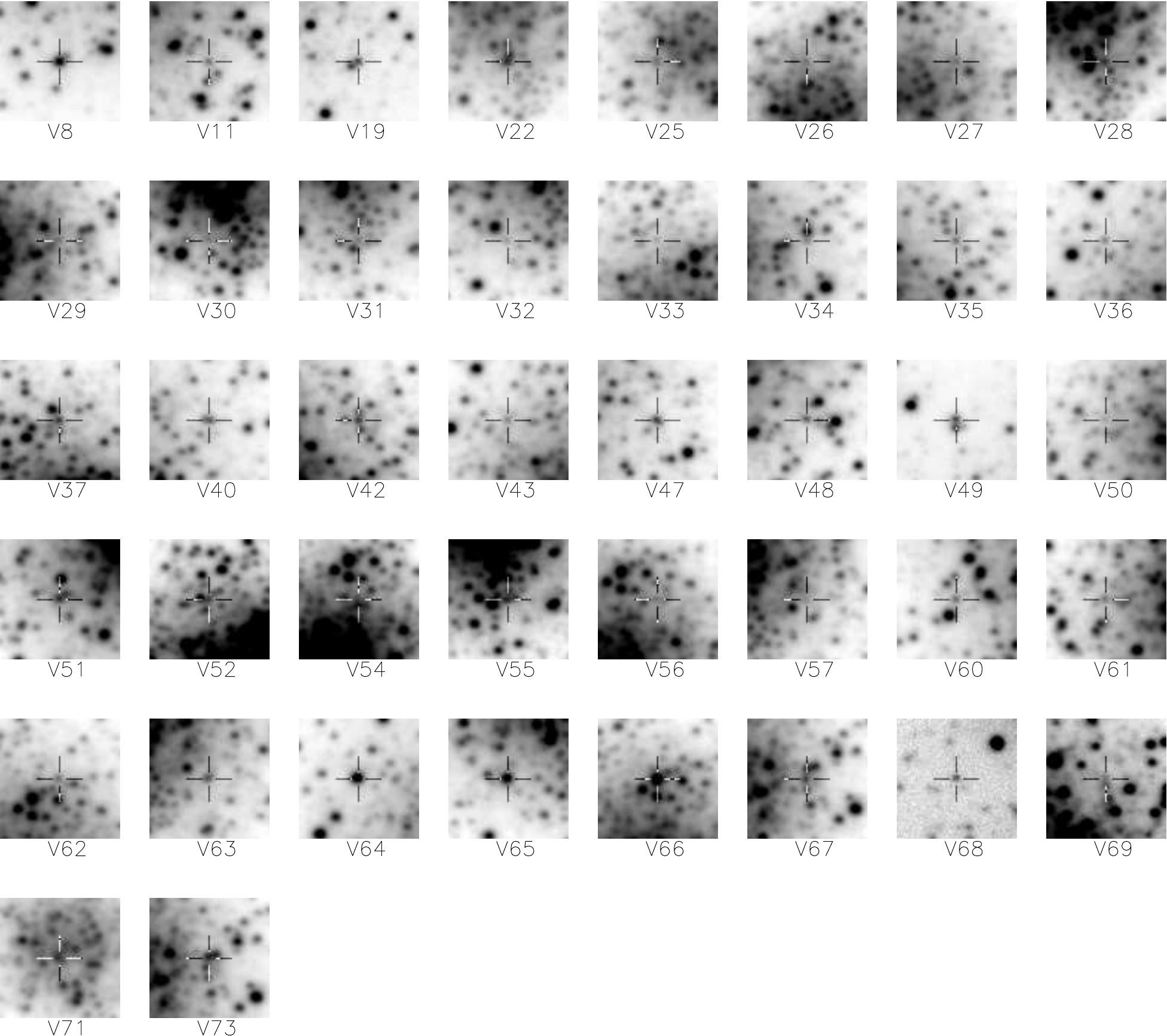}
\caption{Finding chart of selected individual stars. Each stamp
is 23.7$\times$23.7~arcsec$^{2}$.}
\label{stamps}
\end{figure*}

The confirmed variables, their mean magnitudes, amplitudes, periods, times of maximum
light and equatorial coordinates are listed in Table~\ref{variables}. Detailed
finding charts of all variables are given in Figs. \ref{chart} and \ref{stamps}. The
light curves of RRL stars, phased with the ephemerides given in Table~\ref{variables}
are displayed in Figs.\ref{VARSabA} and \ref{VARSc}.  Comments on
individual stars are made in Appendix A.

From the above methods of variable star finding we estimate that our search is
complete down to about $V \sim 18.5$ for amplitudes larger than 0.03 mag and
periods between 0.2 d and a few tens of days in the outer regions of the cluster.
However,
we note that the core is very concentrated and our subtractions there are noisy. 
Therefore, our variable search in the core is not complete. In two recent papers
Skottfelt et al. (2013; 2015) have shown the benefits of using electron multiplying
CCD (EMCCD) photometry and DIA to extract faint variables in the dense central
regions of globular clusters. This is achieved from a high frame-rate time-series 
and by shifting and adding frames to counteract the blurring atmospheric effects which
contributes towards a high spatial resolution. Thus we believe the core of
NGC~6229 would be an excellent target for EMCCD photometry.

\begin{figure*} 
\includegraphics[width=15.0cm,height=22.cm]{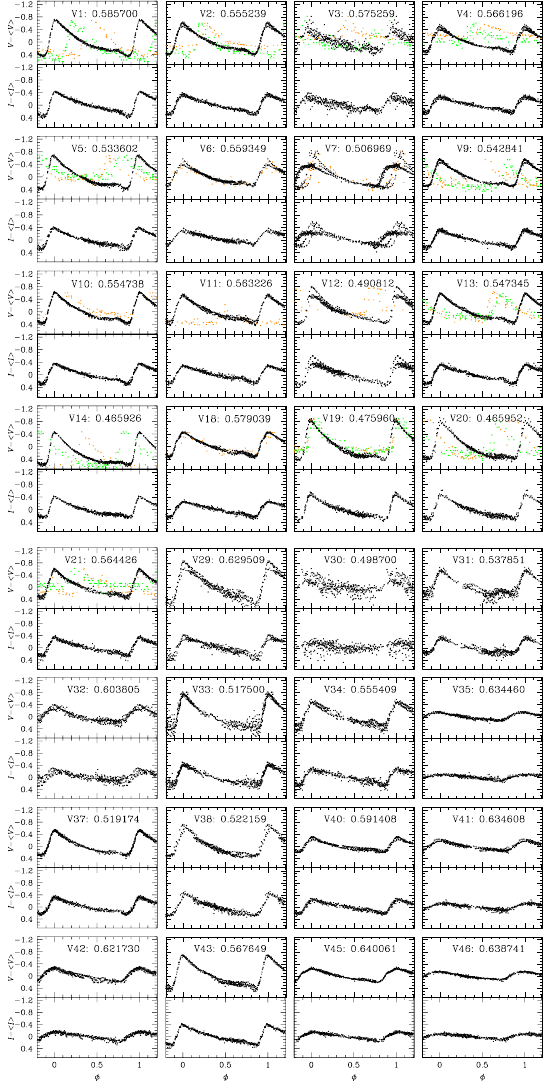}
\caption{Light curves of the RRab stars in NGC~6229. 
They are phased with the ephemerides listed in Table~\ref{variables},  the periods
in days are given next to the identification. 
To preserve the vertical scale and limits the vertical axes display 
$V-<V>$ and $I-<I>$; the individual mean values are also listed in
the table.
Orange circles
represent the data from Baade (1945) and green circles are data from
Mannino (1960), whenever available. Phase differences between the archival data and
our data suggest secular period variations that are calculated in $\S$
\ref{sec:Pdot}. Variables with names V49 and onwards are new discoveries in
this paper.}
   \label{VARSabA}
\end{figure*}

\begin{figure*} 
\addtocounter{figure}{-1}
\includegraphics[width=15.cm,height=8.5cm]{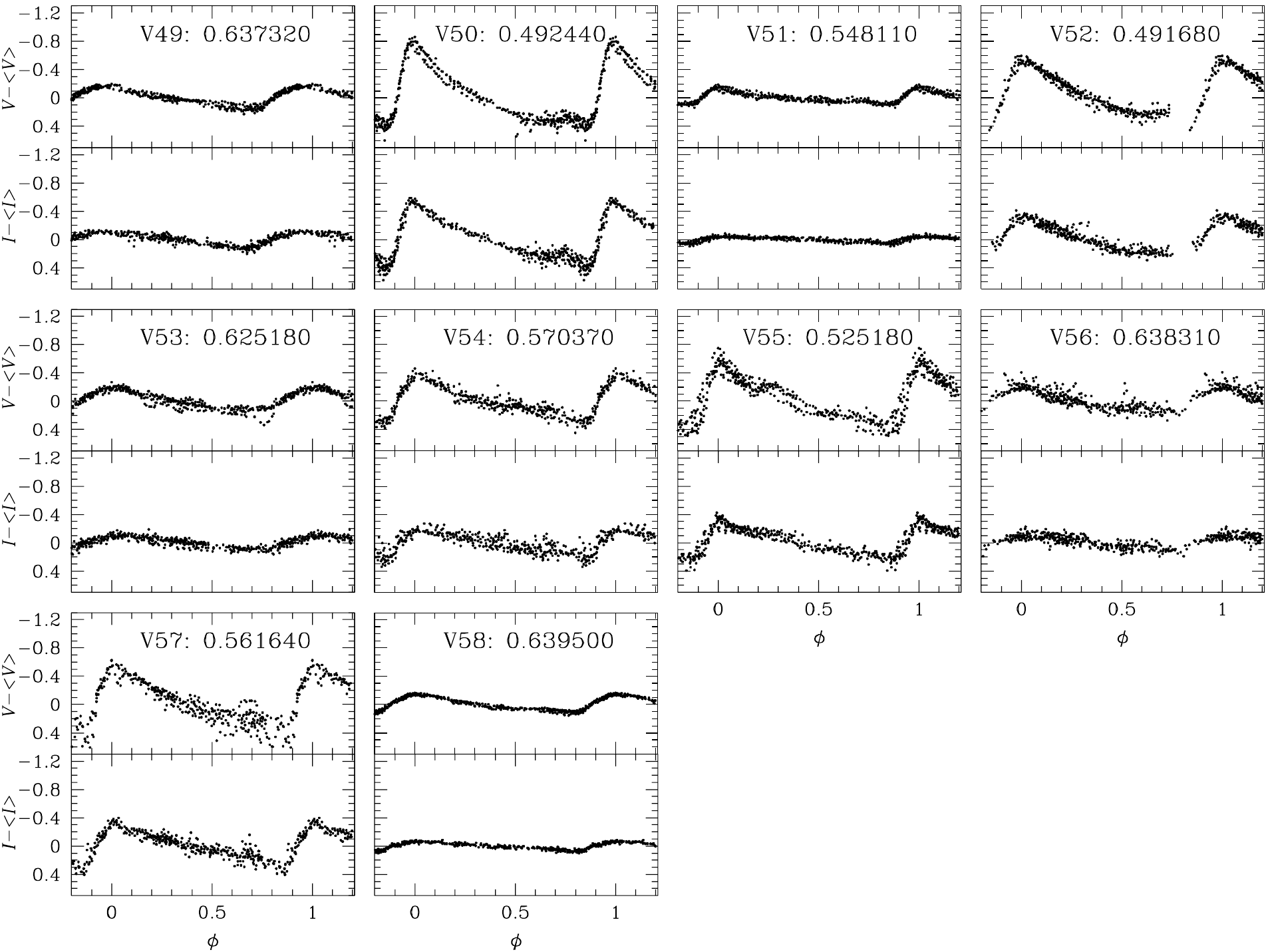}
\caption{Continued}
\end{figure*}

\begin{figure*} 
\includegraphics[width=15.cm,height=11.cm]{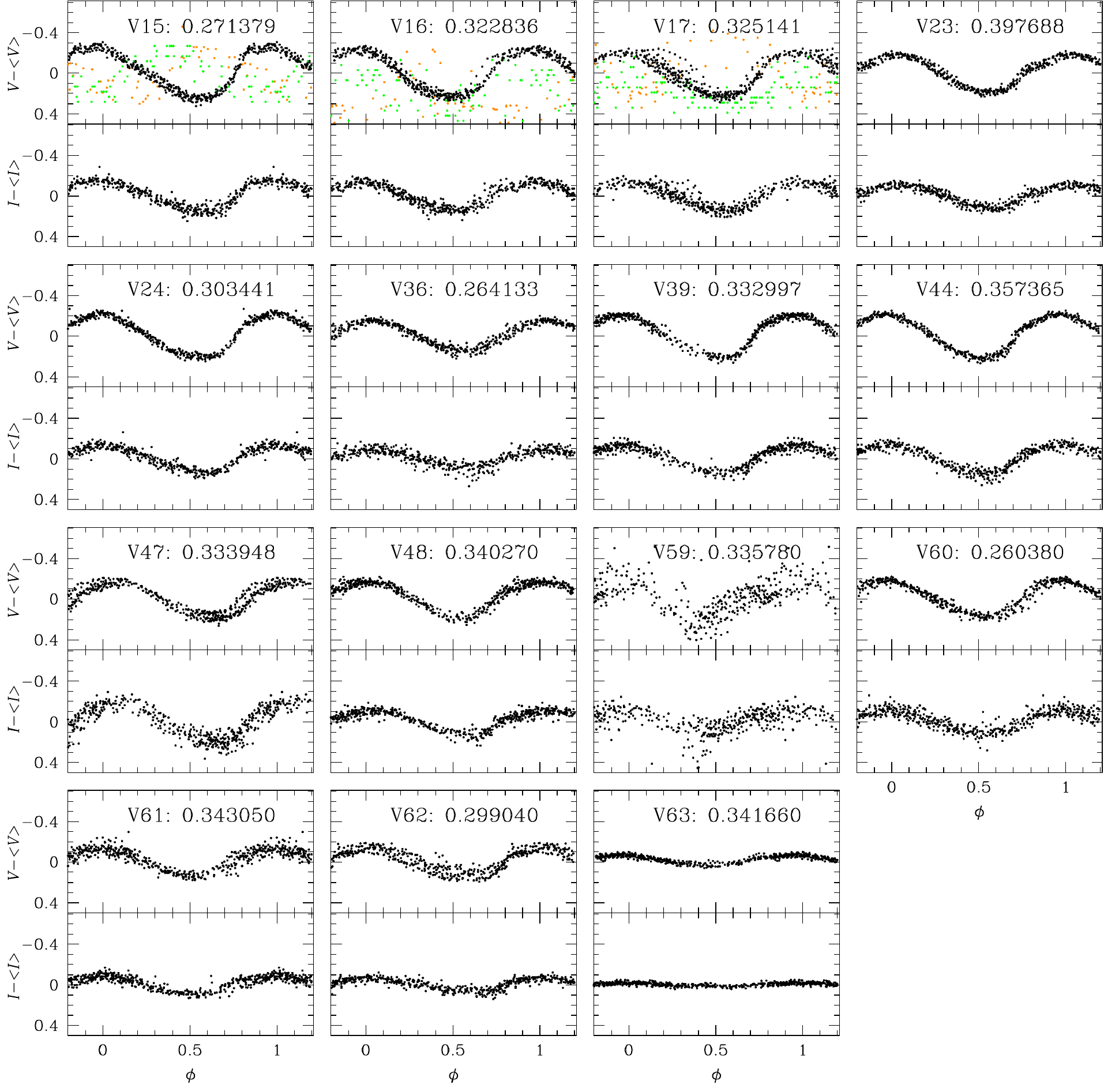}
\caption{Light curves of the RRc stars in NGC~6229
phased with the periods listed in Table~\ref{variables}.}
    \label{VARSc}
\end{figure*}

\begin{figure*} 
\includegraphics[width=17.cm,height=5.cm]{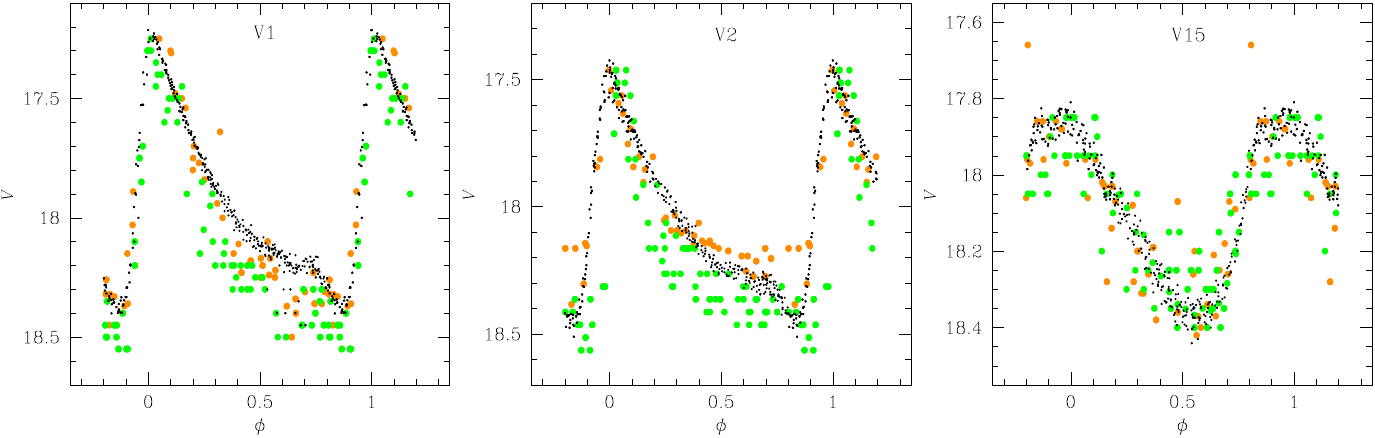}
\caption{Examples of light curves phased with the ephemerides including the period
change rate $\beta$ which are listed in Table~\ref{variables}. Colour symbols are as
in Fig. \ref{VARSabA}.}
    \label{Pdot}
\end{figure*}

\subsection{RR Lyrae stars with a secular period variation}
\label{sec:Pdot}

The combination of our data with those from Baade (1945) and Mannino (1960),
available for variables V1-V21, give a
time-base of about 81 years. For the benefit of future work on these variables
we have uploaded these archival data into the Strasbourg astronomical Data Center
(CDS)\footnote{data are available at the following links:
http://cdsarc.u-strasbg.fr/cgi-bin/VizieR?-source=J/ApJ/102/17  and
http://cdsarc.u-strasbg.fr/cgi-bin/VizieR?-source=J/other/MmSAI/31.187}. For these
stars the light curves of the archival data
phased with the ephemerides in Table \ref{variables} are also shown in Figs.
\ref{VARSabA}
and \ref{VARSc}. Prominent phase shifts
are visible for the different data sets in several variables, suggesting secular
period
variations. Despite the large time-base, the number of times of maximum  that
can accurately be determined for each variable is limited, preventing us from
calculating the secular period change with the classical O-C residuals analysis.
Alternatively, we used the approach introduced by Kains et al. (2015) for the
case of NGC~4590, which for completeness is described below. The period change can be
represented as 

\begin{equation}
 P(t)=P_0+\beta(t-E),
\end{equation}

\noindent
where $\beta$ is the period change rate  expressed in units of d d$^{-1}$, and $P_0$
is the period at the epoch $E$. The number of cycles at time $t$
elapsed since the epoch $E$ is:

\begin{equation}
\label{eq:NE}
N_E = \int_{E}^{t} \frac{dx}{P(x)} = \frac{1}{\beta} {\rm ln}[ 1 +
\frac{\beta}{P_0}(t-E)],
\end{equation}

\noindent
hence the phase at time $t$ can be calculated as

\begin{equation}
\label{eq:chanceperiod}
 \phi(t)=N_E-{\left \lfloor N_E \right \rfloor}.
\end{equation}

\begin{figure} 
\includegraphics[width=8.cm,height=13.cm]{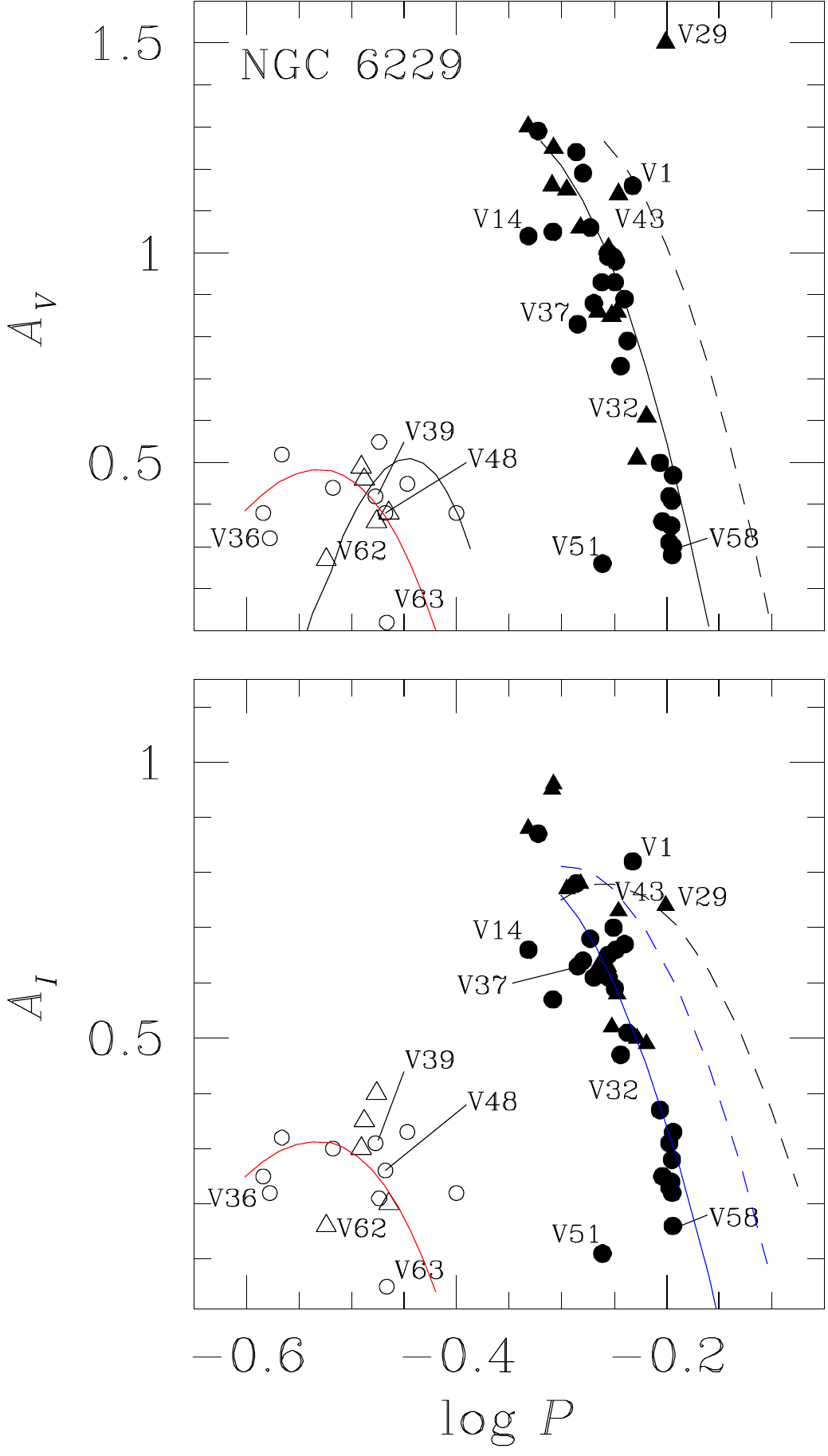}
\caption{The log P vs. amplitude plane for the $V$ and $I$
amplitudes of the RRL in NGC~6229.
Filled and open symbols represent RRab and RRc stars respectively. Triangles are used
for stars with Blazhko modulations. In the top panel the continuous and segmented
lines
are the loci found by Cacciari et al. (2005) for unevolved and evolved RRab stars,
respectively, in the OoI cluster M3. The black parabola was
found by
Kunder et al. (2013b) from 14 Oo II clusters. We calculated the red parabola from a
sample of RRc stars in five Oo I clusters and avoiding Blazhko variables. In the
bottom panel the black segmented locus
was found by Arellano Ferro et al. (2011, 2013) for the OoII
clusters NGC~5024 and NGC~6333. The blue loci are from
Kunder et al. (2013a) for the OoI cluster NGC~2808. The red
parabola was calculated from RRc stars in three Oo I clusters with $I$ amplitudes.}
    \label{fig:Bailey}
\end{figure}

Then we construct a grid ($P$,$\beta$) with values in steps of $10^{-6}d$ and 0.004
d Myr$^{-1}$ in $P$ and $\beta$ respectively, around a given starting period $P_0$, 
and examine the light curve dispersions for each pair. We choose as a solution the
one that produces the minimum
dispersion of our data combined with those of Baade (1945) and Mannino (1960). In
Fig. \ref{Pdot} we show three examples of the light curves phased with eq.
\ref{eq:chanceperiod}, those of the RRab stars V1 and V2 and the RRc V15. The
resulting $P$ and $\beta$ for those stars with an evident
period change are listed in Table \ref{variables}. In a few cases where the archival
data are very scattered, we have not been able to find a reliable solution (e.g.
for stars V17 and V18); however, their secular variation is clear and the proper
calculation will have to wait until sufficient data become available.

The period change rates that we found are very scattered; they average $-0.116 \pm
0.201$ d Myr$^{-1}$. While this number is consistent with zero, the scatter is
very large. 
Taking the sign of the period change rate as an indication of the
direction in which the star is evolving, we could not discern for NGC~6229 a
systematic direction towards which stars are evolving. A summary of the observed
period changes in RR Lyrae stars in globular clusters is offered by
Catelan (2009), in particular his Fig. 15, according to which the period change rate
$\beta$ is related to the ${\cal L} = {\rm (B-R)/(B+V+R)}$ Lee-Zinn parameter, defined
to
describe the stellar distribution on the HB (Zinn 1986; Lee 1990),
where B, V, R refer to the number of stars to the blue, inside and to the red of the
RR Lyrae instability strip. Catelan's diagram shows that, for clusters with
${\cal L} < 0.5$ the value of $\beta$ is negligible, and it increases exponentially
for larger ${\cal L}$, i.e. for clusters with blue HB's. From star counts in our CMD
we estimate for
NGC 6229 ${\cal L} = 0.18$, which  can be compared with the value 0.24 reported by
Mackey
\& van den Bergh (2005), and that would imply $\beta \sim 0.$ Despite this
the average $\beta$ found in this paper is very negative but with a large scatter.
This may imply that rather erratic period changes dominate the period changes
produced otherwise by stellar evolution across the instability strip in NGC~6229.

\subsection{Bailey diagram and Oosterhoff type}

The periods of the 42 RRab and 15 RRc stars listed in Table \ref{variables} average 
0.561d and 0.322d, respectively. These values clearly indicate that NGC~6229 is an
Oosterhoff type I cluster.

The Bailey diagram for the $V$ and $I$ filters is shown in Fig. \ref{fig:Bailey}. The
distribution of the RRab stars in each plane is in close agreement with the loci
identified for Oo I clusters by Cacciari et al. (2005) (M3), Kunder et al. (2013a)
(NGC~2808) and Arellano Ferro et al. (2014) (NGC~3201). A number of outliers are
noted: the case of V51 can be understood since its amplitude is uncertain
due to its Blazhko nature. For V14 and V37, with clean non-modulated light curves,
we do not have an explanation. In the upper end we note V1 is the only
non-modulated star sitting near the
locus for evolved stars, however its position on the HB (see the bottom panel of
Fig. \ref{CMD}) does not add to that interpretation. 
The Bailey diagram further confirms NGC~6229 as an Oo I type cluster.

The distribution of RRc stars is expected to have a parabolic form (Bono et al.
1997) as in fact it has been found in some Oo II clusters  
e.g. NGC 2419 (Di Criscienzo et al. 2011), M9 (Arellano Ferro et al. 2013), M55 (Olech
et al. 1999), M30 (Kains et al. 2013) and Kunder et al. (2013b) for a group of 14 Oo
II
clusters. 
We have assembled a sample RRc stars in Oo I clusters (M3, M5, M79, NGC~6229 and
NGC~6934) with determined amplitudes in $V$ and/or $I$ and without obvious
amplitude modulatios. From their distribution (not shown here) we calculated the
red parabolas shown in Fig. \ref{fig:Bailey}. The scatter about these parabolas is
sometimes considerable, like in the case of Oo II clusters of Kunder et al. (2013a).
The mathematical expresions for these parabolas are:

\begin{equation}
A_V=-3.95 + 30.17 P - 51.35 P^2,
\end{equation}

\begin{equation}
A_I=-2.72 + 20.78 P - 35.51 P^2.
\end{equation}

\section{RR Lyrae stars: [Fe/H] and M$_V$ from light curve Fourier decomposition}
\label{sec:RRLstars}
 
Stellar physical parameters, such as [Fe/H], M$_V$, $T_{\rm eff}$, 
mass and radius for RR Lyrae stars can be calculated via the
Fourier decomposition of their $V$ light curves into its harmonics as:

\begin{equation}
m(t) = A_0 ~+~ \sum_{k=1}^{N}{A_k ~\cos~( {2\pi \over P}~k~(t-E) ~+~ \phi_k ) },
\label{eq_foufit}
\end{equation}

\noindent
where $m(t)$ are magnitudes at time $t$, $P$ is the period and $E$ the epoch. A linear
minimization routine is used to derive the amplitudes $A_k$ and phases $\phi_k$ of
each harmonic, from which the Fourier
parameters $\phi_{ij} = j\phi_{i} - i\phi_{j}$ and $R_{ij} = A_{i}/A_{j}$ are
calculated. The mean
magnitudes $A_0$, and the Fourier light curve fitting parameters of
individual RRab and RRc stars are listed in Table
~\ref{tab:fourier_coeffs}. In this table we have excluded stars with evident
amplitude modulations or excessive noise.

\begin{table*}
\caption{Fourier coefficients $A_{k}$ for $k=0,1,2,3,4$, and phases $\phi_{21}$,
$\phi_{31}$ and $\phi_{41}$, for RRab and RRc stars. The numbers in parentheses
indicate
the uncertainty on the last decimal place. Also listed is the deviation 
parameter $D_{\mbox{\scriptsize m}}$ (see Section~\ref{sec:RRLstars}).}
\centering                   
\begin{tabular}{lllllllllr}
\hline
Variable ID     & $A_{0}$    & $A_{1}$   & $A_{2}$   & $A_{3}$   & $A_{4}$   &
$\phi_{21}$ & $\phi_{31}$ & $\phi_{41}$ 
&  $D_{\mbox{\scriptsize m}}$ \\
     & ($V$ mag)  & ($V$ mag)  &  ($V$ mag) & ($V$ mag)& ($V$ mag) & & & & \\
\hline
       &       &   &   &   & RRab stars    & &        
   &             &       \\
\hline
V1 & 17.936(1)& 0.383(2)& 0.206(2)& 0.132(2)& 0.088(2)& 3.988(12)& 8.316(18)&
6.422(26) &  0.9 \\
V5 & 18.160(2)& 0.359(2)& 0.174(2)& 0.128(2)& 0.081(2)& 3.876(18)& 8.099(27)&
6.153(37) &  2.0 \\
V10 & 18.101(2)& 0.332(2)& 0.158(2)& 0.120(2)& 0.075(2)& 3.894(18)& 8.198(25)&
6.232(38) & 1.0 \\
V11 & 18.119(3)& 0.302(3)& 0.151(3)& 0.102(2)& 0.065(3)& 4.043(29)& 8.386(43)&
6.526(63) &  0.4 \\
V13 & 18.106(2)& 0.308(3)& 0.159(3)& 0.112(2)& 0.069(2)& 3.886(25)& 8.238(36)&
6.235(53) & 1.4 \\
V14 & 17.840(2)& 0.366(3)& 0.182(3)& 0.125(3)& 0.078(3)& 3.737(22)& 7.790(32)&
5.681(48) &  1.9 \\
V18 & 18.066(2)& 0.273(2)& 0.128(2)& 0.091(2)& 0.052(2)& 4.124(24)& 8.420(36)&
6.823(56) & 3.7 \\
V19 & 18.130(2)& 0.432(3)& 0.209(3)& 0.153(3)& 0.101(3)& 3.811(20)& 7.957(29)&
5.906(42) & 1.5 \\
V21 & 18.063(2)& 0.304(3)& 0.152(3)& 0.105(3)& 0.070(3)& 4.011(32)& 8.293(47)&
6.308(95) & 1.5 \\
V35 & 18.062(1)& 0.115(2)& 0.039(2)& 0.020(2)& 0.002(2)& 4.259(64)& 8.924(111)&
7.049(84) & 6.0 \\
V37 & 17.947(2)& 0.282(3)& 0.138(3)& 0.095(3)& 0.057(3)& 3.959(29)& 8.171(43)&
6.237(63) & 1.1 \\
V45 & 18.027(1)& 0.174(2)& 0.073(2)& 0.037(2)& 0.014(2)& 4.190(31)& 8.794(55)&
7.540(13) & 26.9 \\
V46& 17.965(1)& 0.115(2)& 0.039(2)& 0.018(2)& 0.009(2)& 4.280(31)& 9.042(55)&
7.899(13) & 32.5 \\
V51 & 16.658(1)& 0.068(2)& 0.040(2)& 0.031(2)& 0.022(2)& 4.050(71)& 8.312(101)&
6.436(183) & 4.1 \\
V58 & 17.319(1)& 0.102(1)& 0.046(1)& 0.023(1)& 0.007(1)& 4.046(14)& 8.664(14)&
7.056(13) & 4.1 \\
\hline
             &            &           &           &           & RRc stars &           
 &             &             &\\
\hline
V15 &18.118(2)& 0.252(2)& 0.056(2)& 0.015(2)& 0.014(2)& 4.635(43)& 2.716(158)&
1.449(167) & --\\
V23 &17.868(1)& 0.185(2)& 0.009(2)& 0.011(2)& 0.007(2)& 5.742(211)& 4.718(183)&
3.143(270) & --\\
V24 &18.016(1)& 0.217(2)& 0.030(2)& 0.015(2)& 0.014(2)& 4.522(64)& 3.262(124)&
2.189(134) & --\\
V36 &17.942(2)& 0.149(2)& 0.005(2)& 0.002(2)& 0.003(2)& 4.609(464)& 4.956(1296)&
3.515(755) & --\\
V39 &18.113(2)& 0.210(2)& 0.032(2)& 0.018(2)& 0.011(2)& 4.695(71)& 3.834(127)&
2.525(209) & --\\
V44 &17.988(1)& 0.216(2)& 0.027(2)& 0.008(2)& 0.010(2)& 4.883(73)& 4.051(246)&
2.750(185) & --\\
V48 & 17.849(2)& 0.182(3)& 0.019(2)& 0.009(2)& 0.006(2)& 5.726(130)& 4.425(269)&
1.978(420) & -- \\
V60 &18.016(2)& 0.176(2)& 0.027(3)& 0.005(2)& 0.003(3)& 4.402(95)& 2.649(481)&
0.893(751) & --\\
V63 &16.750(1)& 0.046(1)& 0.002(1)& 0.004(1)& 0.004(1)& 4.550(89)& 4.105(328)&
2.100(362) & --\\
\hline
\hline
\end{tabular}
\label{tab:fourier_coeffs}
\end{table*}

These Fourier parameters and the semi-empirical calibrations of Jurcsik
\& Kov\'acs (1996), for RRab stars, and Morgan et al. (2007),
for RRc stars, are used to obtain [Fe/H]$_{\rm ZW}$ on the Zinn \& West (1984) 
metallicity scale  which have been transformed to the UVES scale using the
equation [Fe/H]$_{\rm UVES}$= $-0.413$ +0.130[Fe/H]$_{\rm ZW}-0.356$[Fe/H]$_{\rm
ZW}^2$ (Carretta et al. 2009).
The absolute magnitude $M_V$ can be derived from the calibrations of Kov\'acs \&
Walker
(2001) for RRab stars and of Kov\'acs (1998) for the RRc stars;  these two 
calibrations are naturally based on different sets of calibrators in assorted
Galactic and LMC globular clusters and are therefore independent. However, their zero
points have been tied to the LMC distance modulus of 18.5$\pm$0.1 mag
(e.g. Clementini et al. 2003) as discussed in detail by Arellano Ferro et al. (2010).
The effective temperature $T_{\rm eff}$ was estimated using the calibration of Jurcsik
(1998). For brevity we
do not explicitly present here the above mentioned calibrations; however, the
corresponding equations, and most
importantly their zero points, have been discussed in detail in previous papers
(e.g. Arellano Ferro et al. 2011; 2013) and the interested reader is referred to them.

It is pertinent to recall that the calibration for [Fe/H] for RRab stars of Jurcsik
\& Kov\'acs (1996) is applicable to RRab stars with a {\it deviation parameter}
$D_m$,
defined by Jurcsik \& Kov\'acs (1996) and Kov\'acs \& Kanbur (1998), not exceeding an
upper limit. These authors suggest $D_m \leq 3.0$. The $D_m$ is listed in
column 10 of Table~\ref{tab:fourier_coeffs}. However we have relaxed the criterion
to include three stars with 3.7 $\leq D_m \leq$ 4.1.  Those RR Lyrae that fall off
the HB, discussed in $\S$  \ref{sec:IND_STARS}, are not included for the luminosity
calculation as indicated in the footnote of Table \ref{fisicos}.

 It is noteworthy at this stage that there is a new calibration of [Fe/H]$_{\rm
UVES}$ for RRc stars, calculated for an extended number of stars and clusters by
Morgan (2013). As described in her paper the results of the new calibration are in
general consistent with the old calibration from Morgan et al. (2007), and that the
average differences are 0.02 dex. We have also carried out the calculation
using the new calibration and found [Fe/H]$_{\rm UVES}=-1.37$, i.e. 0.08 dex more
deficient than the value from the calibration of Morgan et al. (2007), which is in
better agreement with the result from the RRab calibration.

\begin{table*}
\footnotesize
\begin{center}
\caption[] {\small Physical parameters for the RRab and RRc stars. The numbers in
parentheses indicate the uncertainty on the last 
decimal place.}
\label{fisicos}
\hspace{0.01cm}
 \begin{tabular}{lcccccc}
\hline 
Star&[Fe/H]$_{\rm UVES}$ & $M_V$ & log~$T_{\rm eff}$  & log$(L/{\rm L_{\odot}})$ &
$M/{\rm M_{\odot}}$&$R/{\rm R_{\odot}}$\\
\hline
 &  &  & RRab stars &  & & \\
\hline
V1  &$-1.385(18)$& 0.511(3) & 3.809(7) & 1.696(1) & 0.70(6) & 5.69(1) \\
V5  &$-1.394(27)$& 0.611(3) & 3.812(8) & 1.656(1) & 0.70(7) & 5.36(1) \\ 
V10 &$-1.378(25)$& 0.604(3) & 3.810(8) & 1.658(1) & 0.68(7) & 5.42(1) \\ 
V11 &$-1.215(39)$& 0.612(4) & 3.811(10) & 1.655(2) & 0.66(8) & 5.40(2) \\
V13 &$-1.302(34)$& 0.636(4) & 3.811(9) & 1.645(2) & 0.66(7) & 5.32(1) \\
V14 &$-1.437(32)$& 0.711(4) & 3.817(9) & 1.616(2) & 0.71(7) & 5.00(1) \\
V18 &$-1.245(33)$& 0.614(3) & 3.806(9) & 1.655(1) & 0.66(7) & 5.50(1)\\
V19 &$-1.297(27)$& 0.640(4) & 3.820(8) & 1.644(2) & 0.73(7) & 5.10(1) \\
V21 &$-1.317(45)$& 0.610(4) & 3.810(13) & 1.656(2) & 0.66(10) & 5.42(2) \\
V35$^{a}$ &--& 0.664(3) & 3.804(12) & 1.634(1) & 0.56(8) & 5.43(1) \\
V37 &$-1.254(40)$& 0.696(4) & 3.813(10) & 1.622(2) & 0.65(8) & 5.15(2) \\
V45$^{a}$ &--& 0.602(3) & 3.799(7) & 1.659(1) & 0.63(5) & 5.72(1) \\
V46$^{a}$ &--& 0.657(3) & 3.800(21)& 1.637(1) & 0.58(15)& 5.55(1) \\ 
V51$^{b}$ &$-1.229(92)$& -- & 3.804(21)& -- & --&-- \\
V58$^{b}$ &$-1.243(13)$& -- & 3.797(7 )& -- & -- &-- \\

\hline
Weighted &&&&&&\\
Mean & $-$1.31(1)& 0.621(1) & 3.810(2) & 1.652(1) & 0.66(2) &5.43(1)\\
\hline 
 &  &  & RRc stars &  & & \\
\hline
V15 &$-1.22(25)$& 0.603(2) & 3.872(1) & 1.659(1) & 0.61(1) & 4.09(1)\\
V23 &$-1.49(47)$& 0.464(9) & 3.859(1) & 1.714(4) & 0.47(1) & 4.62(2)\\
V24 &$-1.33(23)$& 0.577(3) & 3.868(1) & 1.669(1) & 0.56(1) & 4.21(1)\\
V36$^{a}$  & --       & 0.660(21)& 3.887(8) & 1.636(8) & 0.50(5) & 3.72(4)\\
V39 &$-1.34(26)$& 0.555(3) & 3.865(1) & 1.678(1) & 0.52(1) & 4.31(1)\\
V44 &$-1.50(58)$& 0.527(3) & 3.862(1) & 1.689(1) & 0.50(1) & 4.43(1)\\
V48 &$-1.03(46)$& 0.524(6) & 3.867(2) & 1.690(2) & 0.51(1) & 4.33(1)\\
V60 &$-1.11(69)$& 0.673(4) & 3.874(3) & 1.631(2) & 0.58(2) & 3.92(1)\\
V63$^{b}$ &$-1.27(66)$& --& 3.865(2)& --& -- &-- \\

\hline
Weighted&&&&&\\
Mean & $-$1.29(12) & 0.581(1) & 3.867(1) & 1.668(1) &0.54(1) & 4.18(1) \\
\hline
\hline
\end{tabular}
\end{center}
\raggedright
\center{\quad $^{a}$: No [Fe/H] estimate since $D_m > 4.1$. V36 has a very peculiar
value of $\phi_{31}$; $^{b}$: No $M_V$ estimate since star falls off the HB.}
\end{table*}

The physical parameters for the RR Lyrae stars and the inverse-variance
weighted means are reported in Table~\ref{fisicos}.
Two independent estimations of [Fe/H]$_{\rm UVES}$ have been found from 12 RRab and 8
RRc
stars in NGC~6229: $-1.31\pm0.01$ and $-1.29\pm0.12$ respectively. 

The weighted mean $M_V$ values for the RRab and RRc stars are 0.621$\pm$0.001 mag
and
0.581$\pm$0.001 mag respectively and will be used in section
\ref{sec:distance} to estimate the mean distance to the parent cluster.

\section{On the structure of the Horizontal Branch}
\label{sec:HB}
In this section we analyse the distribution of RRL stars in the HB. 
 The expanded HB in Fig. \ref{CMD} displays in detail the positions of
most of the
RR Lyrae stars. RRab and RRc are distinguished as well as Blazhko and stable
variables. There is a neat segregation between stable RRab and RRc with a border at
$(V-I)_0\sim 0.45$. Similarly clear segregations have been found before in NGC~5024
(Arellano Ferro et al. 2011; their Fig. 4) and NGC 4590 (Kains et al. 2015; their
Fig. 11) both at $(V-I)_0\sim 0.45-0.46$. In the bottom panel of Fig. \ref{CMD} the
corresponding border lines for NGC~5024 and NGC~4590 duly reddened are indicated.
It can be seen that the border line between stable RRab and RRc is the same
in the three clusters.
We should emphasise that the above empirical border is trespassed by a few Blazhko
variables (NGC 6229) and the double mode RRL (NGC 4590). Looking for  this
segregation effect in clusters previously studied by our group, we checked the CMDs
and
noted that with very few exceptions the effect
is also visible in NGC~5053 (Arellano Ferro et al. 2010; Fig. 8)  (except for V10 with
a poor I light
curve), NGC~1904 (Kains et al. 2012) and NGC~7099 (Kains et al. 2013), despite their
low number of RRL. On the other
hand the RRab-RRc splitting is definitely not seen in NGC 3201 (Arellano Ferro et
al. 2014).

\begin{figure} 
\includegraphics[width=8.3cm,height=5.5cm]{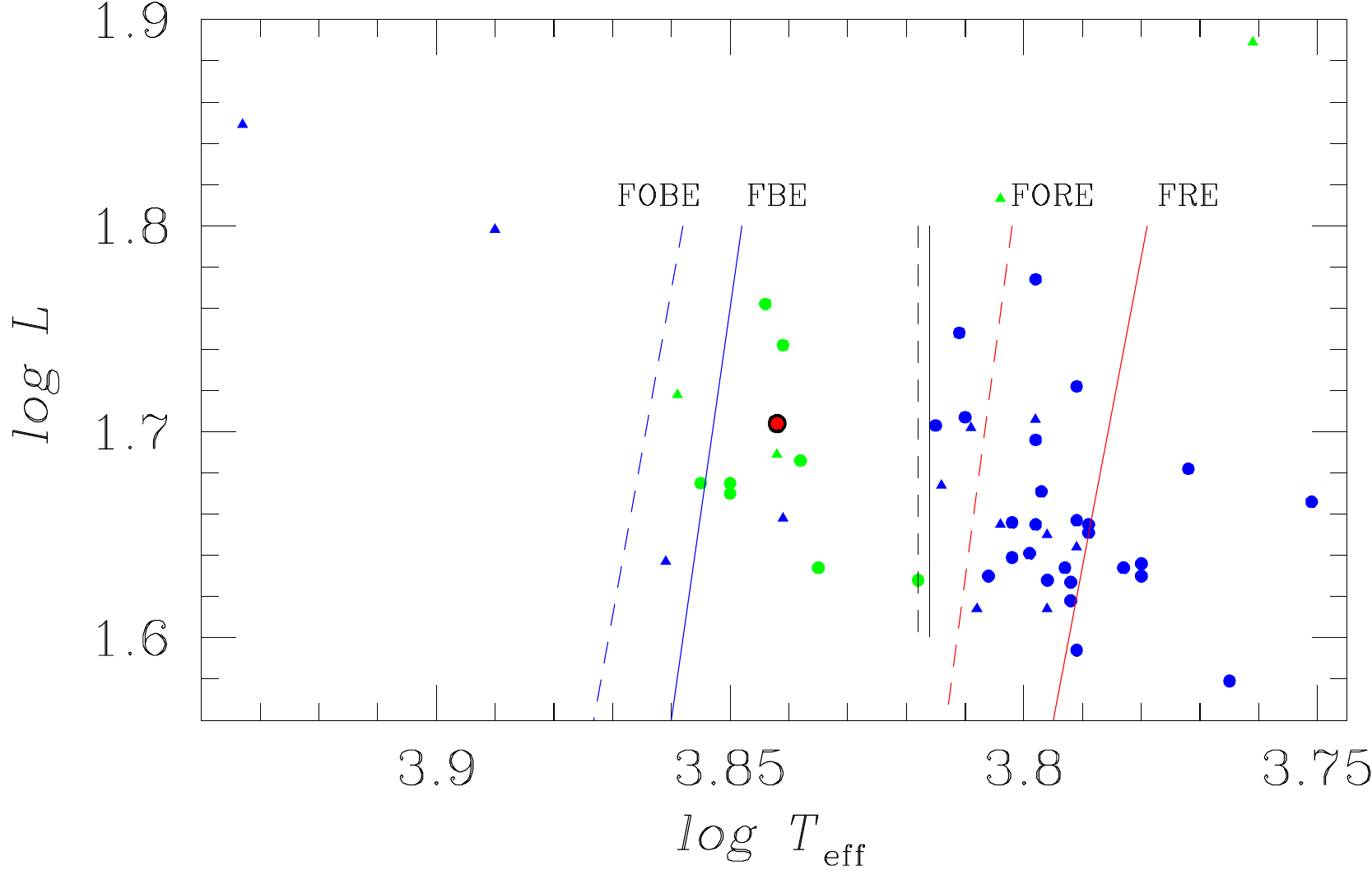}
\caption{The HR diagram of the horizontal branch of NGC~6229. The coloured lines are
the IS borders for the fundamental mode (FM) and the first overtone (FO) according to
Nemec et al. (2011). The two vertical black lines correspond to the empirical border
between
RRab and RRc stars seen in NGC 5024 (solid) and NGC~4590 (dashed). See also 
Fig. \ref{CMD} and the text for a discussion.}
    \label{TLplane}
\end{figure}

It is illustrative to display the HB in the $T_{\rm eff}$-${\rm log} L$ plane. In
order to
convert the observational HB shown in Fig. \ref{CMD} into the
HR diagram, we have proceeded as in Arellano Ferro et al. (2010)
where a polynomial for $T_{\rm eff}$-$(V-I)_0$ is given based on the HB models of
VandenBerg et al. (2006) with the colour-$T_{\rm eff}$ relations as described
by VandenBerg \& Clem (2003). Fig. \ref{TLplane} illustrates the distribution of
RRL using the same colour code as in  Fig. \ref{CMD}. The red and blue
borders of the instability strip (IS) for the fundamental mode (F) and first 
overtone (FO), calculated
by Nemec et al. (2011) based on the Warsaw pulsation code for a mass of 0.65
$M_{\odot}$, are
shown in the figure. The two vertical black lines correspond to the empirical border
between RRab and RRc stars seen in NGC 5024 (solid) and NGC~4590 (dashed) which match
the evident border in NGC~6229 and that, in our opinion, represent the empirical 
red border of the first overtone instability strip (FORE). This is some 100-200K
hotter than the theoretical border found by Nemec et al. (2011). Another feature is
that
the fundamental red edge (FRE) in NGC~6229 extends further to the red by at least
400K. 

Regarding the distribution of RRc stars, they are well constrained to the
first overtone
instability strip. The distribution of RRL on the HB observed in NGC~6229 can
be explained by the arguments of Caputo et al. (1978) which involve the occurrence
of a hysteresis mechanism (van Albada \& Baker 1973) for stars crossing the IS.
According to this mechanism the stars in the  "either-or" region can retain the mode
they were pulsating in before entering the region which depends on whether the star is
coming from the blue side as a RRc or from the red side as a RRab. Furthermore,
Caputo et al. (1978) suggest that the original mode in which a RRL is pulsating
depends on the location of the starting point at the Zero Age Horizontal Branch
(ZAHB), which in turn depends on the mass and the chemical composition of the star;
they used these ideas to explain the existence of the two Oosterhoff groups. A
schematic and clear explanation of this can be found in their Fig. 3. In summary
there are two scenarios; 1) the ZAHB point is in the fundamental zone, leading to an
assortment of RRc and RRab in the "either-or" region and a lower value of the
average period for both RRc and RRab and a lower proportion of RRc stars,
hence an OoI cluster, 
and 2) the ZAHB point is bluer than the fundamental zone, in the "either-or" or in the
first overtone regions, then the "either-or" region is populated exclusively by RRc
stars and the RRab stars are to be found only in the fundamental region, producing 
larger averages of periods and a higher proportion of RRc and hence OoII type cluster.

The above scenario seems to explain well the clean segregation of RRc and RRab 
in OoII clusters with evolution across the IS towards the red. In
agreement with this picture we can mention the analysis of Pritzl et al. (2002b).
These authors have shown how in clusters with blue HBs, i.e. large values of $\cal L$,
stars with masses below a critical mass on the ZAHB to the blue of the IS, evolve
redwards and spend sufficient time to contribute to the population of RRL. These
arguments suggest then that clusters with a blue HB morphology are OoII, with
redwards evolution, which in turn favours the segregation between RRc and RRab at the
IS.

We have explored the CMD of several OoII clusters published by our group over the last
few years (e.g. NGC~288, NGC~1904, NGC~5024, NGC~5053, NGC~5466, NGC~6333, NGC~7099)
and noted the clear segregations, with a few exceptions of indiviual stars with noisy
light curves. The one exception is NGC 4590, a clear OoII with a
red HB ($\cal L$=0.17) and evidences, from the secular period changes analysis of 14
RRL, that evolution does not occur in a preferential direction (Kains et al. 2015).
On the other hand the OoI cluster NGC 3201 ($\cal L$=0.08) as expected, shows no signs
of segregation (Arellano Ferro et al.
2014). Likewise NGC~6229  has a low $\cal L$ (0.24) and no preferential
evolution direction from the scattered period change rate found in $\S$
\ref{sec:Pdot}, in spite of this the RRc and the RRab are neatly separated.

A close inspection of the distribution of clusters of all Oosterhoff types on the
[Fe/H]-$\cal L$ plane from the data in Table 3 of Catelan (2009), reveals NGC~4590
as the OoII cluster with the reddest HB with ${\cal L} < 0.4$ and that there are five
OoI clusters (NGC~6266, NGC~6284, NGC~6402, NGC~6558 and NGC~6626) with blue HB's
(${\cal L} > 0.4$). It would be of great interest to explore their RRL distributions
in the IS.

One more point of interest in Fig. \ref{TLplane} is the distribution of Blazhko
variables. We have plotted with triangles only stars that show clear Blazhko
amplitude modulations while marginal cases (e.g. V50, V52, V54) are plotted as stable
stars. In his paper Gillet (2013) concluded that the Blazhko effect is not a
consequence of the transition between modes and that a star that penetrates the
"either
or" region (RRc or RRab) becomes a Blazhko variable, as seems to be supported
by the field Blazhko RRL plotted in Gillet's Fig. 1.
The later statement implies that all stars  in the "either or" region should be
Blazhko variables, which is not observationally supported by the distribution of
Blazhko stars in NGC~4590, NGC~5024 and NGC~6229. Some outlier Blazhko variables 
from the IS in NGC~6229, mostly RRab stars, are likely not the result of hysteresis
in the change of mode process as once suggested by Arellano Ferro et al. (2010) but
rather a consequence of multi-shock perturbation at the main pulsation instability
yet to be explained by theory (e.g. Gillet 2013).

In the CMD of Fig. \ref{CMD} we have overplotted two versions of the ZAHB for
[Fe/H]=$-1.42$ and [$\alpha$/Fe]=+0.4; for Y=0.25 and Y=0.27, calculated from the 
Victoria-Regina stellar models of VandenBerg et al. (2014). The model for Y=0.25
is a reasonable lower envelope to the distribution of RRL. The difference with a
non-diffusive model for [Fe/H]=$-1.31$, [$\alpha$/Fe]=+0.3 and Y=0.25 from
VandenBerg et al. (2006) is negligible. Apparently the change in the primordial
helium content largely compensates for the difference between diffusive and
non-diffusive models (VandenBerg private communication). The only two stable
RRab's
a little below this ZAHB are V33, which shows some scatter near minimum, and V55 whose
flux is likely contaminated by its neighbour V31 (see the discussion in
$\S$ \ref{sec:RRL}). If the intrinsic thickness of the stellar distribution on the HB
is
between 0.1-0.2 mag, e.g. the case of M3 in Fig. 1 of Cacciari et al. (2005), it is
clear that a number of RRL in NGC 6229 are overluminous. A natural explanation would
be evolution off the ZAHB, however we did not find evidence of stars evolved in
the amplitude-period diagram of Fig. \ref{fig:Bailey}. Also no evidence of systematic 
secular period increases in this rather red-HB cluster have been found (see $\S$
\ref{sec:Pdot}). Furthermore, stars in an advanced evolution towards the AGB should be
brighter and have longer periods, however we find no correlation between $<V>$ and
period for the RRL in NGC~6229.

\begin{figure} 
\includegraphics[width=8.3cm,height=5.5cm]{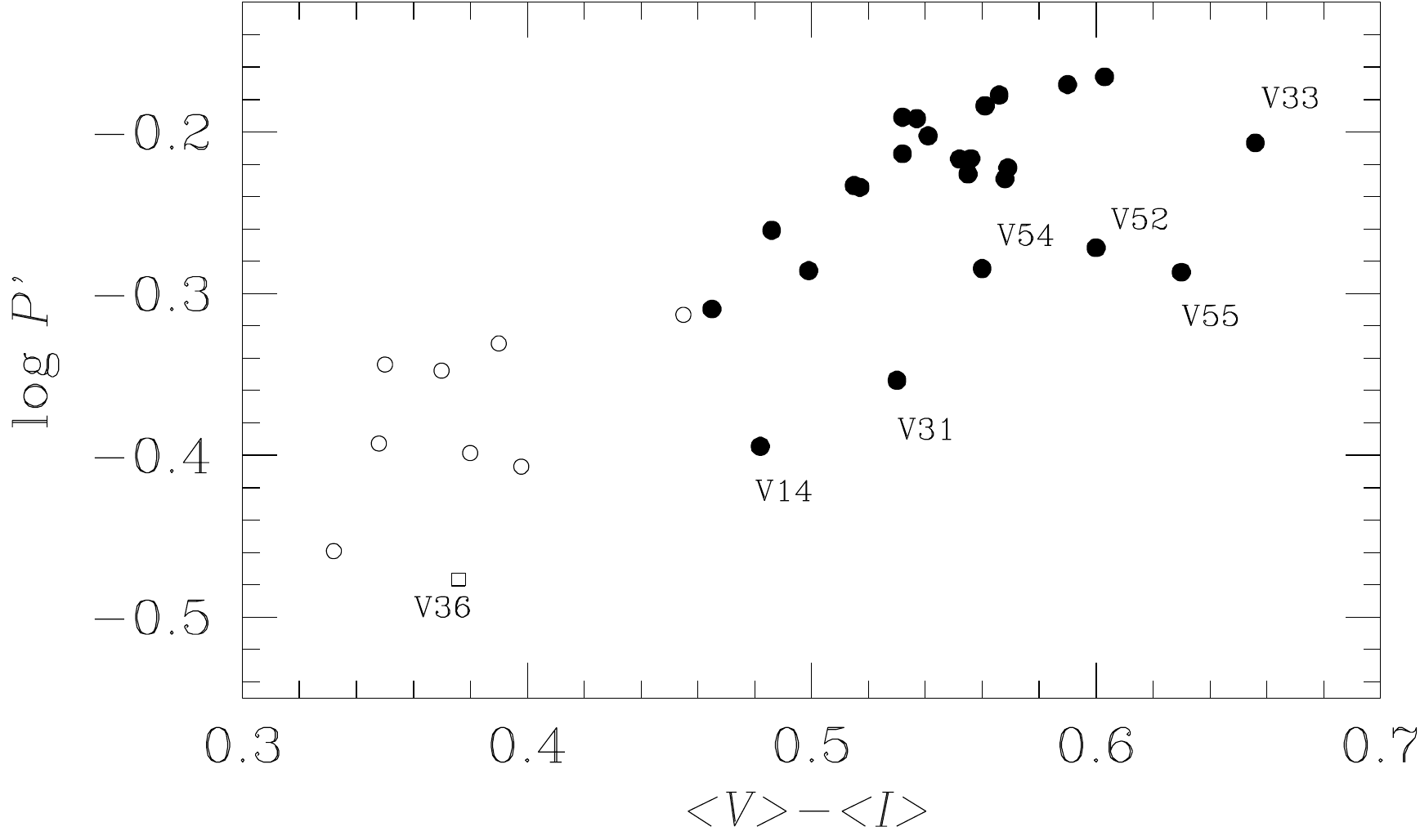}
\caption{Period-colour diagram of stable RRL stars. Filled circles are used for
RRab stars and empty circles for RRc stars. The periods of the RRc stars were
fundamentalized by adding 0.127 to $\log P$. The reduced period $P'$ is defined
as $\log P'= \log P + 0.336 (<V> -18.021)$. See text for further discussion. 
}
    \label{VI_logP}
\end{figure}

There may be other reasons for overluminosity than
evolution off the ZAHB. For example, helium abundance dispersion
in HB stars could be produced if more than one generation of stars
are mixed in a given GC. D'Antona \& Caloi (2008) have argued that
second-generation stars, formed from enriched material of
first-generation progenitors, may have a spread in Y above the
primordial value, and that these stars will occupy a bluer position
on the HB than the first-generation ones, which sit to the red
of the instability strip. It has been noted by Catelan et al.
(2009) that HB stars with higher Y should also be brigther due to their
stronger H-burning shells, and they have searched for such an effect in M3.
By comparing with HB models with Y in the range 0.23-0.28, they
concluded that no systematic overluminosity in blue HB stars is evident,
and hence the Y spread in M3 should not be larger than 0.01.
For NGC 6229 we have overplotted ZAHB models from VandenBerg (2014) for
Y = 0.25, 0.27, and 0.285, and found no evidence of the blue HB stars
being brighter than the red ones, so no spread in the value of Y is
immediately obvious in this cluster either. Similar tight
constraints would apply to the case of helium mixing on the RGB 
phase (Sweigart 1997).

Another reason may be the presence of an unseen companion. In this context it is of
interest to explore the period-colour relation for stable RRL of Fig. \ref {VI_logP}.
In this diagram the period  ${\rm log} P'= {\rm log} P + 0.336 (<V> -18.021)$
where 18.021 mag is the average $<V>$
for all the stable RRL. The concept of ${\rm log} P'$  was introduced by van Albada \&
Baker
(1971) and it was later slightly modified by Bingham et al. (1984). It significantly
reduces the scatter in the period-colour diagram and is
useful for identifying stars with peculiar photometric properties such as colours too
red
or too blue, and luminosities too bright or too faint relative to what is expected
from
their periods. In Fig. \ref {VI_logP} we note some outliers and have labelled them.
V33 has been commented on above, while V36 was believed to be an RRd in the CVSGC
but classified here as an RRc with a very short period among RRc stars. It is worth
noting that V14, V31, V54 and V55 are also the brightest stable RRab in the HB (see
the bottom panel of Fig. \ref{CMD}). Their position on Fig. \ref {VI_logP} implies
that they have too short a period for their colour, which implies that they have
larger gravity and
should be of lower luminosity. The fact that instead they are among the brightest
in the HB, implies that they have an extra source of luminosity and this may be an
unseen companion. It was through the same lines of argumentation that
Cacciari et al. (2005) seem to have identified three RRL in M3 (V48, V58 and V146)
that may have companions.

Thus it seems likely that not only evolution has contributed to the large
dispersion of RRL in the HB of NGC~6229 but also a dispersion in the primordial helium
abundance and in some cases the presence of unseen companions.

\section{NGC~6229 distance and metallicity from its variable stars}
\label{sec:distance}

Four independent estimates of the distance to the cluster are possible from our data;
from the 
weighted mean $M_V$, calculated for the RRab and RRc from the Fourier 
light curve decomposition (Table \ref{fisicos}),
which can be considered as independent since they come from different empirical
calibrations;
from the one known SX Phe star and the P-L relation; and from the tip of the RGB
bolometric magnitude calibration. Below we expand on these solutions.
 
 Since NGC~6229 is a high Galactic latitude object and despite its large distance,
the interstellar extinction and reddening are negligible, we adopted
$E(B-V)$=0.01 (Harris 1996). Given the mean $M_V$ for RRL in Table \ref{fisicos} we
found a true distance modulus of
$17.385\pm0.117$ mag and $17.433\pm 0.096$ mag using the RRab and RRc stars
respectively, which correspond to the distances $30.0\pm 1.5$ and $30.7\pm 1.1$ kpc.
The quoted uncertainties are the standard deviation of the mean.
The distance to NGC~6229 listed in the catalogue of Harris (1996) (2010 edition) is
30.5 kpc in good agreement with our calculations.

A third independent estimate of the distance to the cluster is via the one SX Phe
identified in this work, V68, assuming that its period 0.0384563d corresponds to the
fundamental mode. We have used the P-L relation of Cohen \&
Sarajedini (2012);

\begin{equation}
M_V = -(1.640 \pm 0.110) - (3.389 \pm 0.090) {\rm log} P_f
\label{eqn:PL_CS}
\end{equation}

\noindent
where $P_f$ refers to the fundamental mode given in days. The above equation 
implies a true distance modulus of 17.303$\pm$0.168 or a distance of $28.9\pm
2.2$ kpc. The uncertainty was calculated from calibration errors neglecting the
uncertainty of the period. It is worth noting that eq. \ref{eqn:PL_CS}, when
applied to a sample of $\delta$ Scuti stars in the LMC produces a distance modulus of
18.49$\pm$0.10 (Cohen \& Sarajedini 2012) which is consistent with the modulus
18.5$\pm$0.1 used to set the zero points of the absolute magnitude calibrations for
RRab and RRc stars (see Arellano Ferro et al. 2010 $\S$ 4.2). Thus the distances
derived here for the RRL and the SX Phe are on a consistent distance scale.

Yet another approach to the cluster distance determination is by using the tip of
the
RGB. This method was developed for estimating distances to nearby galaxies (Lee,
Freedman \& Madore, 1993). Here we used the calibration of Salaris \& Cassisi
(1997) for the bolometric magnitude of the tip of the RGB in terms of the overall
metallicity: 

\begin{equation}
M_{bol}^{tip} =-3.949 -0.178[M/{\rm H}]+0.008[M/{\rm H}]^2
\label{eqn:Mbol}
\end{equation}

\noindent
where $[M/\rm H]= {\rm[Fe/H]}+\rm log(0.638f +0.362)$ and log f = [$\alpha$/Fe]
(Salaris et al. 1993).

We found that the method is extremely sensitive to the star selection as the
tip of the RGB, hence we restricted our calculation to the five reddest stars 
in the red box in the CMD of 
Fig. \ref{CMD}. We have assumed [$\alpha$/Fe]=+0.4 (e.g Jimenez \& Padoan 1998)
and found an average distance of $34.9\pm2.4$ kpc.

Within their own uncertainties the above distance determinations are satisfactorily
in agreement. However, since the SX Phe determination is based on only one star 
and the tip of the RGB approach is very sensitive to the selected RGB tip stars,
which in turn is scattered,
we believe that the distance determination from the RRL stars is the most
accurate.

The average value of [Fe/H]$_{\rm UVES}=-1.31\pm0.12$ found from the Fourier
decomposition of the light curves of 12 RRab and 8 RRc stars in NGC~6229 compares
very well with the
numerous determinations in the literature. In the present edition of his
catalogue of globular clusters Harris (1996) lists an average [Fe/H]$_{\rm ZW}=-1.47$
(or [Fe/H]$_{\rm UVES}=-1.37$) for NGC~6229. A list of
individual results from an assortment of indicators can be found in Table 7 of
Borissova et al. (1999);  they range from $-1.26$ to $-1.44$ in the ZW
scale or from $-1.14$ to $-1.34$ in the UVES scale with uncertainties running within
0.08-0.15 dex.

\subsection{A brief comment on the age of NGC 6229}
\label{sec:age}

We have not attempted an independent estimation of the age of NGC~6229 since
a thorough effort based on deep photometry has
already been carried out by Borissova et al.
(1999). In that paper the authors soundly argue from their differential age
analysis that NGC 6229 and M5 are coeval 
within 1 Gyr. To the best of our knowledge no direct estimation of the absolute age
of NGC~6229 has been carried out. However, the age of M5 was determined by 
Jimenez \& Padoan (1998) via its luminosity function and they found an age of  
$10.6\pm0.8$ Gyr taking into account [$\alpha$/Fe]=+0.4. 
Two more recent determinations of the age of M5 are by Dotter et al. (2010) using
relative ages isochrone fitting and by VandenBerg et al.
(2013) using an improved calibration of the "vertical method" or the magnitude
difference between the turn off point and the HB; these authors find 12.25$\pm$0.75
Gyr and 11.50$\pm$0.25 Gyr respectively.
Thus, in the CMD of Fig.
\ref{CMD} we have overlayed the corresponding isochrones for 
12.0 Gyr from the Victoria-Regina evolutionary models (VandenBerg et
al. 2014)\footnote{http://www.canfar.phys.uvic.ca/vosui/\#/VRmodels} 
and for the metallicities [Fe/H]=$-1.42$ (red) and [Fe/H]=$-1.31$
(blue) and [$\alpha$/Fe]=+0.4, shifted for the average apparent distance modulus
$\mu$=17.440 mag found from the RRab and RRc stars, and $E(V-I)=1.616~E(B-V)=0.016$. 

The isochrones for these two metalicities are very similar except perhaps at the
tip of the RGB where the richer isochrone gets less luminous. However, if a vertical
shift is applied to these isochrones within the uncertainty of the distance modulus,
0.01 mag, then the two cases are indistinguishable.
We can stress then that our CMD is consistent with the metallicity and
distance derived in this paper and an age of 12.0$\pm$1.0 Gyr found in the above
papers for the coeval cluster M5.

\section{Summary of results and conclusions}
We have significantly updated the census of variable stars in NGC 6229. The DIA
approach to the
CCD image reduction and analysis allowed us to recover all of the 48 variables
previously known and to identify 25 new variables; 10 RRab, 5
RRc, 6 SR, 1 CW, 1 SX Phe and two new variables that we were unable to classify.

Numerous new Blazhko variables have been identified among both the already known and
the new RRL stars, including five RRc variables (V16, V17, V47, V61 and V62) (see
Table
\ref{variables}). Also our data do not support the Blazhko
effect seen in V5, V18, V33, V34, V35, V37, V46 and V48 by BCV01
based on the $D_m$ parameter which
is very sensitive to light curve coverage and scatter. 

Thanks to the inclusion of archive data from 1932-1935 (Baade 1945) and from
1956-1958 (Mannino (1960) we have noted secular period variations in 16 RRL and
have calculated the period change rate in 13 of them.

The variability of V29 and V30 is confirmed and they are classified here as RRab. On
the other hand the variability of previously suspected variables V25, V26, V28 is not
confirmed. The classifications of V31, V32 and V39, previously believed to be EB, RRc
and RRab respectively, are now corrected as RRab, RRab and RRc.

The metallicity and distance of NGC~6229 were calculated via the light curve Fourier
decomposition of RRab and RRc stars to find averages of [Fe/H]$_{\rm UVES}=-1.31 \pm
0.12$
and $d=30.3\pm 1.4$ kpc. These results agree with previous independent determinations
and the observed CMD matches the above results and the isochrone of age
$12.0\pm1.0$ Gyr (Jimenez \& Padoan 1998; Dotter et al. 2010; VandenBerg et al.
2013).

Regarding the RRL distribution in the HB we have stressed once more the clear
splitting between stable RRc and RRab with an empirical border at about $(V-I)_0=
0.45-0.46$ mag, clearly observed in NGC~6229 and equally clear in NGC~4590 and
NGC~5024, despite being of different Oosterhoff type. This border is interpreted as
the red edge of the first overtone instability strip. Blazhko variables and double
mode stars do not obey that border.

\section*{Acknowledgments}

We are indebted with Prof. Peter Stetson for providing us with unpublished data for 
standard stars in the field of NGC~6229, and to Prof. Don VandenBerg for providing
ZAHB sequences based on the Victoria-Regina stellar models from 2014 and for very
enlightening comments. The numerous comments and suggestions from an anonymous
referee are warmly appreciated. This publication was made possible by grants
IN104612-14 and IN106615-17 from the DGAPA-UNAM, Mexico and by NPRP grant \#
X-019-1-006 from the National Research Fund (a member of Qatar Foundation).
We have made an extensive use of the SIMBAD and ADS services, for which we
are thankful.

\appendix

\section[]{Comments on Individual Stars}
\label{sec:IND_STARS}

In this appendix we comment on the light curves, variable types and nature of some 
interesting or peculiar variables in Table \ref{variables}. We put some emphasis
on the amplitude and phase modulations of the Blazhko type  
in specific stars and note that several variables catalogued as Blazhko
variables by BCV01 are not confirmed by our data and that we have detected a
few new ones. In their study, these authors combined older photographic
data converted into $B$ magnitudes with their own $B$ DAOPHOT and ISIS photometry to
evaluate the
amplitude variations (see their Fig. 7 for example). Then they calculated
the {\it deviation parameter} $D_m$, (Jurcsik \& Kov\'acs 1996; Kov\'acs \& Kanbur
1998) and for classification purposes assumed that RRab stars with $D_m > 5$ are
Blazhko variables but note however, that examples exist of Blazhko and non-Blazhko
variables
that do not obey such criterion as $D_m$ which is quite sensitive to the coverage of
the
light curve.  In the present case, all the stars we catalogue as Blazhko variables
are judged from our homogeneous data sets in the
$V$ and $I$ filters, where the amplitude modulations are neatly distinguished in
the light curves in Figs. \ref{VARSabA} and \ref{VARSc}.
Hence, some inconsistencies in the identification of Blazhko variables between the
two studies are expected and we make them explicit in the following paragraphs.

\subsubsection{RR Lyrae stars}
\label{sec:RRL}

{\bf V2, V9, V12, V43.} These stars show Blazhko modulations not identified as such by
BCV01. Stars V2, and V9 also undergo secular period variations which are reported in
Table \ref{variables}.

\noindent
{\bf V3, V59.} These RRL stars display very scattered light curves despite 
being located in an uncrowded region. We find no evidence of either star to be
blended. However, we note relatively poor image subtractions for these stars and in
this area of the field, possibly due to the proximity with the image edge. These are
not Blazhko modulations.

\noindent
{\bf V5, V18, V33, V34, V35, V37, V46, V48.} We find no evidence of amplitude
modulations in these
stars classified as Blazhko variables by BCV01.

\noindent
{\bf V7.} The star was classified as a RRab Blazhko variable by BCV01 based on its
peculiar Fourier coefficients. Our
data show very prominent amplitude modulations ranging about half a magnitude,
but we note that
RRab-like variations are preserved. Our data are insufficient to estimate the Blazhko
period, but an inspection of the amplitudes and dates shows that the first amplitude
decrease took place between April 11th and July 9th 2011. A new amplitude increase
occurred
from March 5th to June 18th, 2013, i.e. in each case the amplitude extrema were
reached within 90 and 100 days.

\noindent
{\bf V16, V17, V47, V61, V62.}
These RRc stars show evidence of amplitude and phase modulations that cannot be
solved  by a secular period change or explained by double mode pulsation. We
consider these
modulations as due to the Blazhko effect.
We note that star V47 falls much to the blue from the HB. However, it is an
unresolved pair (see Fig. \ref{chart}) and undoubtedly its colour has been polluted by
the neighbour.

\noindent
{\bf V29.} It was listed by Spassova \& Borissova (1996) as a probable variable star
and it was not included in the study of BCV01. Our data 
plotted in Fig. \ref{VARSabA} clearly show a RRL type light curve with a small
amplitude modulation. Thus we confirm it to be a RRab variable with the Blazhko
effect. For
its period
(0.629509d) its amplitude in both $V$ and $I$ is peculiarly large, as seen in Fig.
\ref{fig:Bailey}.

\begin{figure*} 
\includegraphics[width=17.cm,height=9.cm]{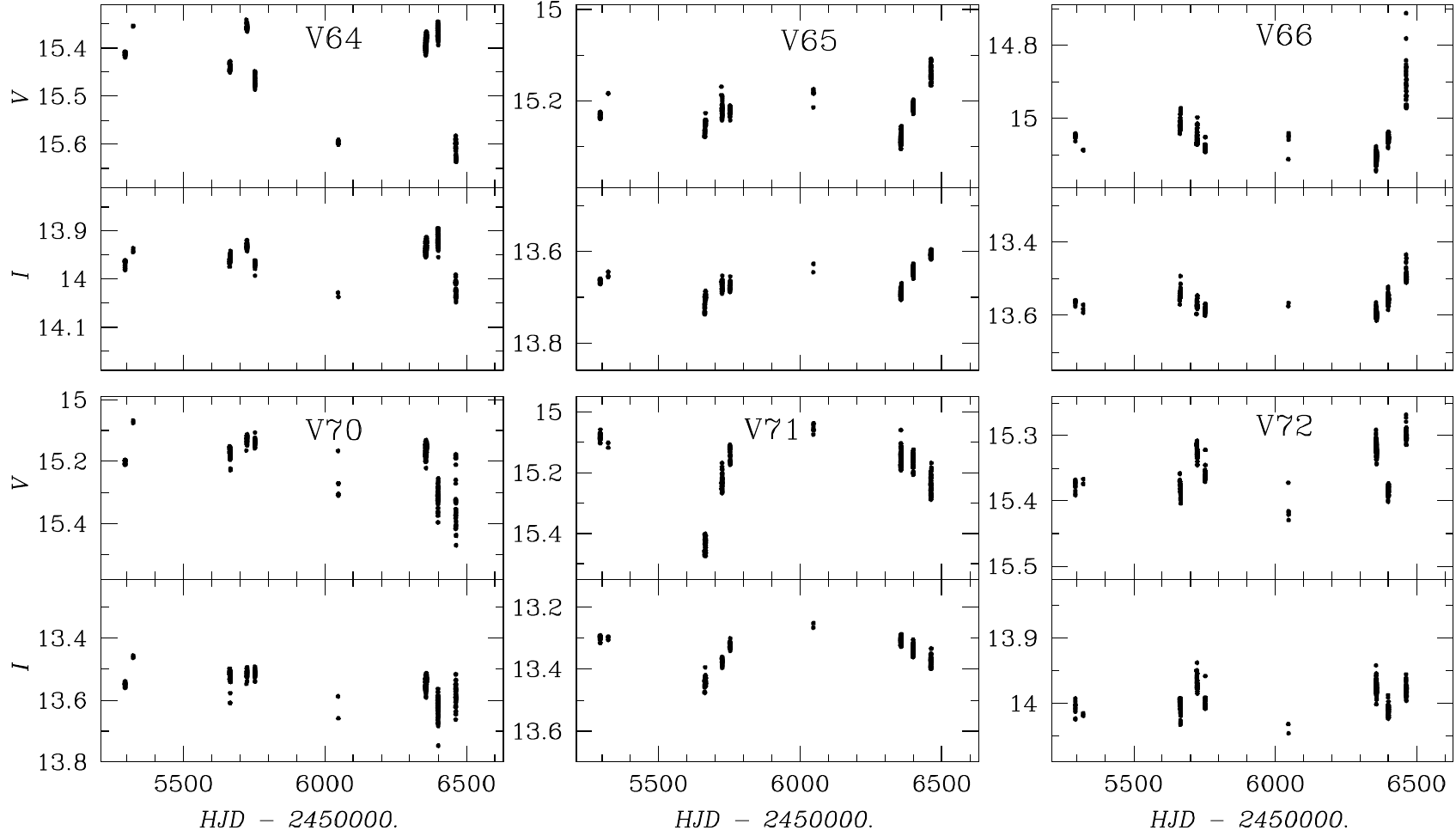}
\caption{Light curves of the six new SR stars.}
    \label{newSRs}
\end{figure*}

\begin{figure*} 
\includegraphics[width=17.cm,height=5.cm]{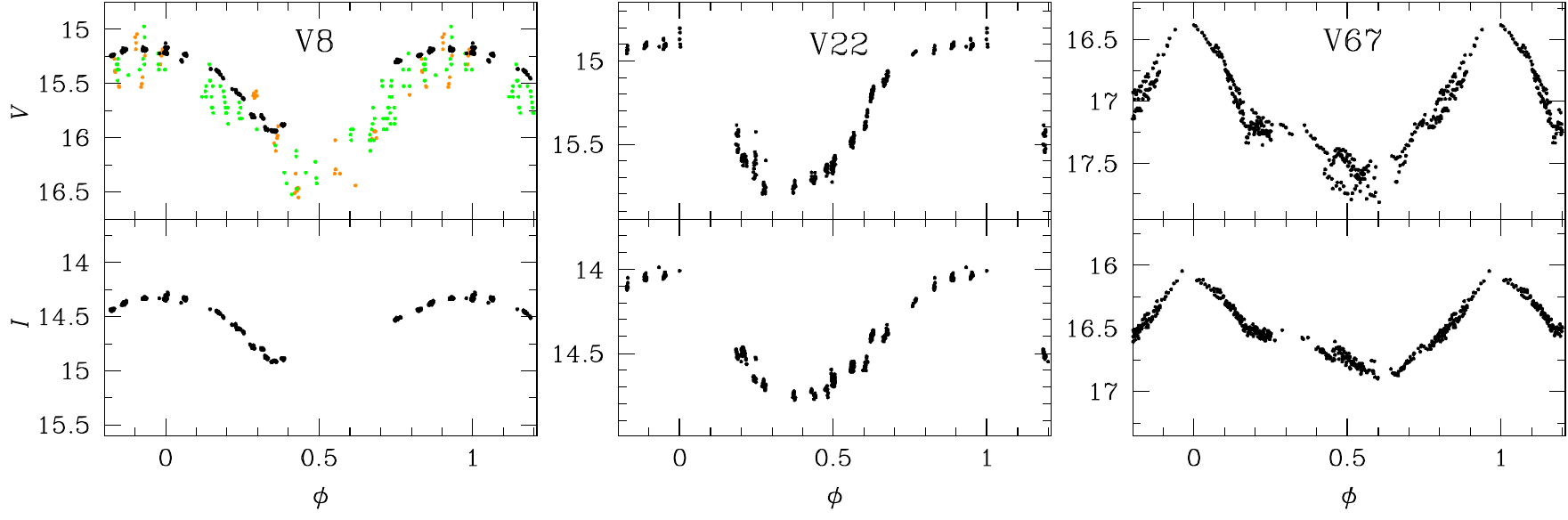}
\caption{The three CW stars in NGC~6229 phased with their periods in Table 3.
For V8 the data from Baade (1945) (orange) and Mannino (1960) (green) 
are included and show the stability of its period. In V67 small 
amplitude modulations are visible.}
    \label{CWs}
\end{figure*}

\begin{figure*} 
\includegraphics[width=17.cm,height=10.cm]{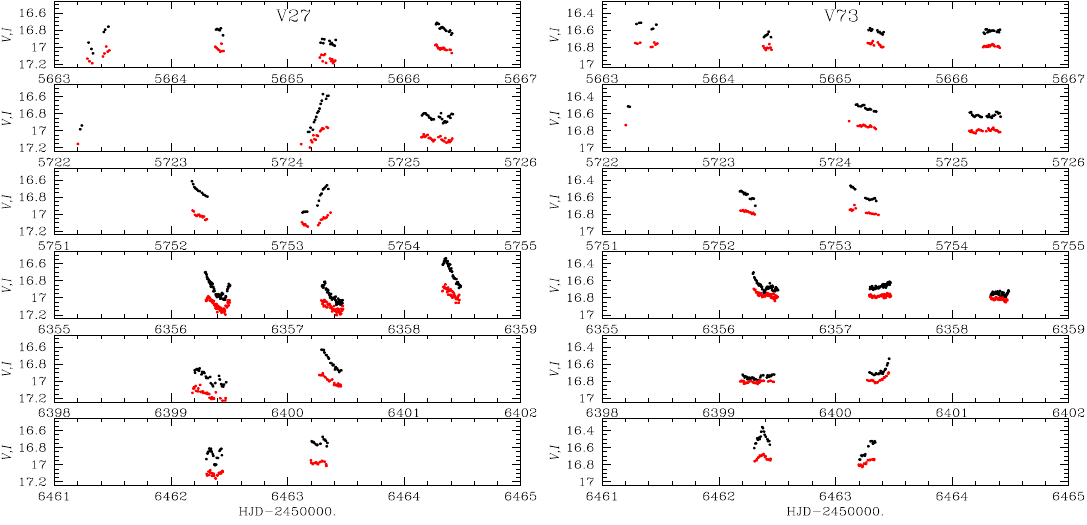}
\caption{Light variations of the possible CW stars V27 and V73. Black and red
circles are the $V$ and $I$ magnitudes respectively. The $I$ magnitudes have been 
arbitrarily shifted to accomodate them in the same box, so that
the variations in both filters can be appreciated. Note that $V$-$I$ variations
are also evident in both stars.}
    \label{fig:V2773}
\end{figure*}

\noindent
{\bf V30.} This is a faint and blended star quite close to the cluster core
displaying a very noisy light curve. Despite this, its light curve
shows a RRab-like shape.
The star was reported as a possible variable by Spassova \& Borissova (1996) and was
not included in the paper by BCV01. Hence, our light curve in Fig. \ref{VARSabA}
is the first one published for this star. We confirm the
variability and suggest that the star is likely a background RRab which would be
consistent with its fainter and redder position on the CMD.

\noindent
{\bf V31 and V55.} V31 was classified by BCV01 as an eclipsing binary with a period
of 0.698893d. Our data do not phase well with that period but instead we find  
0.537851d. The light curve phased with this period appears like that of a RRab
star (Fig. \ref{VARSabA}). 
V31 is very near a new variable star identified in this work,
V55, also a RRab star. The periods of these two stars are similar
but not equal, since each one does not phase the light curve of the
neighbouring star. Also the times of maximum do not coincide. We have confirmed that
both stars are clearly variable by blinking the difference images. The amplitude
modulations visible in both stars however, are most likely not authentic as the
light
curves of both stars may be influencing each other. Nevertheless, we do not confirm
V31
as an eclipsing binary but instead we classify it as RRab. Both V31 and V55 lie well
within the HB and,
in the Bailey diagram (Fig. \ref{fig:Bailey}), they also fall right on the RRab stars
locus.

\noindent
{\bf V32.}
This star is reported as RRc in the CVSGC. We find however that the period 
0.603805d phases the light curve better and produces a RRab shape with amplitude
modulations. The star's position in the Bailey diagram (Fig. \ref{fig:Bailey}) and in
the CMD confirm the above classification.

\noindent
{\bf V35, V41, V45, V46.} The light curves of these RRab call
attention for their small amplitudes which, however, are consistent
with their long periods as can be corroborated in Fig. \ref{fig:Bailey}.

\noindent
{\bf V36.} It is listed as a possible double mode variable in the CVSGC. We have found
no
signs of a secondary frequency in our data but just a period of 0.264133d. 
We classify this star as a RRc and it is labelled accordingly in the CMD and the 
period-amplitude plane.

\noindent
{\bf V39.} This star is listed as RRab in the CVSGC. However, our data favour a
period, light curve shape and position on the Bailey diagram and CMD all consistent 
with the star being a RRc. It must be stressed that the light curve in BCV01
is missing the rise to maximum and the maximum itself, which may have biased their
classification.

\noindent
{\bf V48.} This star has been classified by BCV01 as a RRab with a period of
0.516412d.
Our light curve shows a poor phasing with that period and a glance of the light curve
in BCV01 also reveals a poorly phased sinusoidal curve. We find that a period of 
0.340270d phases the light curve well (Fig. \ref{VARSc}). This period and the
amplitude place the star well within the RRc star domain in the Bailey diagram (Fig.
\ref{fig:Bailey}). Thus we have reclassified the star as RRc.

\noindent
{\bf V50.} This RRab star shows clear amplitude modulations which we attribute to the
Blazhko effect.

\noindent
{\bf V51, V58, V62, V63.} By their periods, light curves morphology and position in
the Bailey diagram of Fig. \ref{fig:Bailey} these are all clear RRL stars. However, in
the CMD they share a common, peculiar position to the red and above the HB. The most
extreme cases are V51 (RRab) and V63 (RRc); both
are unresolved blends (Figs. \ref{chart} and \ref{stamps}) which, at
least partly, explains their
low amplitudes and hence their peculiar position in Fig. \ref{fig:Bailey}. V58 (RRab)
and
V62 (RRc) fall at the expected place on the $A_V-{\rm log} P$ plane and are more
isolated but perhaps not enough for their colours to be unaffected.

\subsubsection{SR stars}
\label{sec:SRs}
{\bf V64-V66, V70-V72.} Their light curves as a function of HJD are displayed in Fig.
\ref{newSRs}. While the variations are clear, given their
likely long characteristic times of variation, our data are insufficient to attempt a
period determination.

\subsubsection{Population II variables}
\label{sec:PII}

\noindent

{\bf V8, V22.} The periods of these two CW stars given in the CVSGC are confirmed in
this work (Table \ref{variables}) and the phased light curves are displayed in Fig.
\ref{CWs}.

{\bf V27, V73.} V27 was classified as a Pop
II cepheid by BCV01 with a period of
1.13827d; they noted however that their light curve was not well defined
and regarded the period as uncertain. Our data
are poorly phased with this period. 
Our data are plotted as a function of HJD in Fig.\ref{fig:V2773} and they display
clear 
variations that are difficult to reconcile with one single period. V73 is 
a newly found variable and, like V27, shows irregular but clear variations. 
Our data seem insufficient for a more detailed analysis of the
characteristic time of variation in these two stars. Their positions on the rms 
and the CMD diagrams of Figs. \ref{rms} and \ref{CMD} seem to support their
classification as
CW stars. Hence we tentatively classify them as such.

\noindent
{\bf V67.} In the CMD this newly discovered variable is found about one
magnitude above the HB and near V27. Its period of 1.57547d and light curve shown
in Fig. \ref{CWs} make it a clear CW variable. It is of interest to comment 
whether or not V67 is an anomalous cepheid. Given the distance modulus of
NGC~6229, its absolute magnitude is $\sim -0.5$. Plotted on the $P-L$ relation
for anomalous cepheids pulsating in the fundamental mode
(see fig 6. of Pritzl et al. 2002a), V67 falls well below the main
relation, 
i.e. it is about 0.5 mag fainter than anomalous cepheids with a 
similar period. Hence V67 is not an anomalous cepheid.

\subsubsection{Variable blue stragglers}
\label{sec:VBS}

\begin{figure} 
\includegraphics[width=8.cm,height=5.cm]{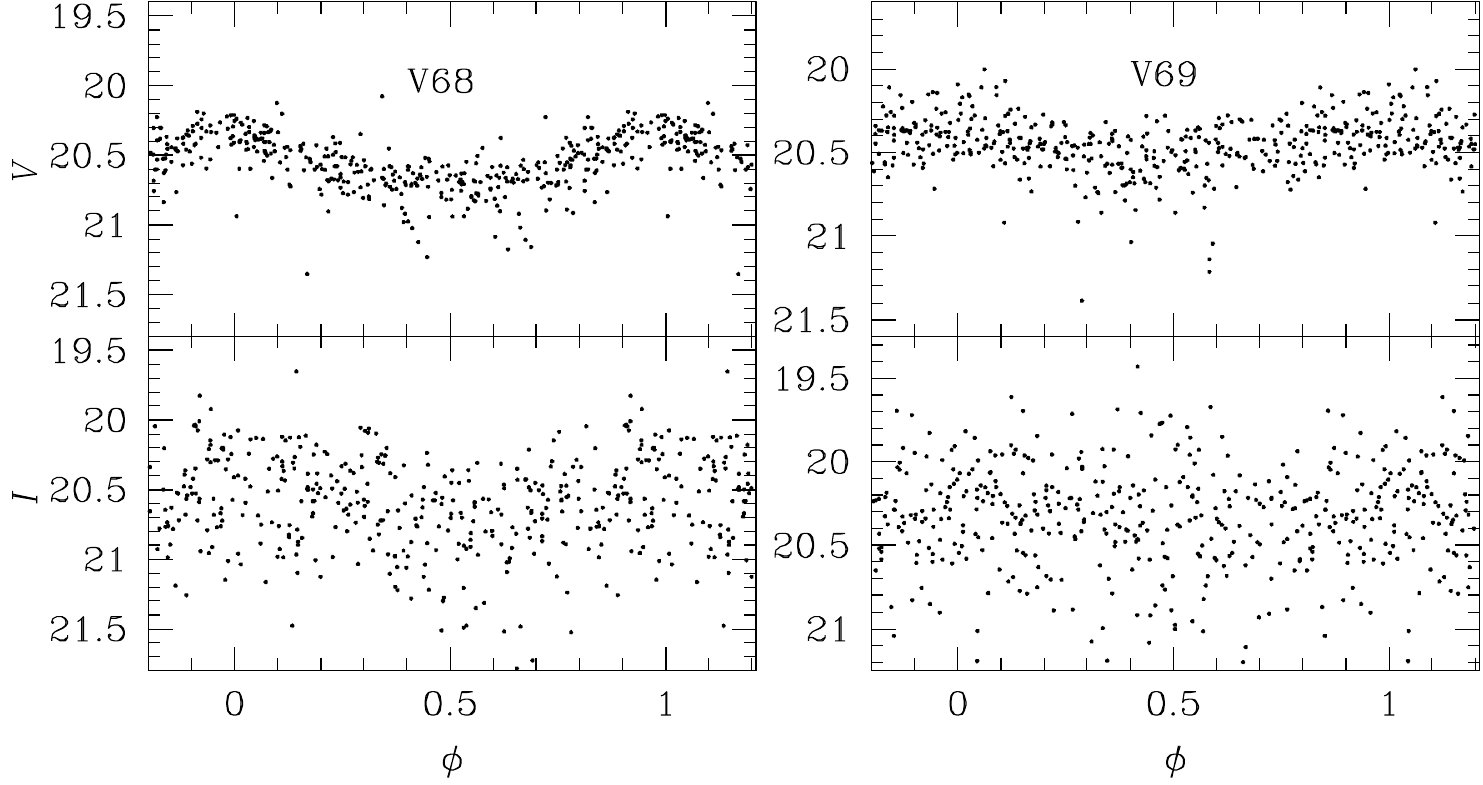}
\caption{Two variables in the blue stragglers region. V68 is an SX Phe star. We have
not been able to classify V69. However, given its period, V69 could
be a background RRc star, see $\S$ \ref{sec:VBS}.}
    \label{SXs}
\end{figure}

{\bf V68 and V69} show variability in our data and are contained in the blue
stragglers region. Their light curves are shown in Fig. \ref{SXs}. 
V68 is a clear SX Phe star for which only one period 
has been found, 0.0384563d. V69 on the contrary displays a small amplitude 
in $V$ and our $I$ data are
too noisy to see the variation. Its period 0.271264d is rather large for an SX Phe
star thus we refrain to classify it as such. Alternatively, the star could be a
background star of a different type, perhaps a RRc.

\subsubsection{Variables not confirmed}
\label{sec:DUDOSAS}

\noindent
{\bf V25, V26, V28.} These stars were listed among a group of twelve possible
variables by  Spassova \& Borissova (1996). The numbers were assigned by BCV01
although they did not include them in their study (as also did not include V29 and
V30). We found no convincing signs of variability. The three stars are very close to
the cluster core and then their light curve RMS values are large;  0.054, 0.035 and
0.071 mag in $V$ and 0.030, 0.024 and 0.036 mag in $I$ respectively. Despite not
being variable, these stars are identified in Figs. \ref{chart} and \ref{stamps}  and
their
photometry included in Table \ref{tab:vi_phot} for the sake of completeness. 

\end{document}